\documentclass{aa}  
\usepackage{enumitem}
\usepackage{graphicx}
\usepackage{cancel}
\usepackage{txfonts}
\usepackage[pdfencoding=auto,psdextra]{hyperref}
\usepackage[dvipsnames]{xcolor}
\hypersetup{
    colorlinks=true,
    linkcolor=Red,
    anchorcolor=Red,
    filecolor=Violet,      
    urlcolor=magenta,
    citecolor=Blue,
}

\begin{document} 

   \title{The Three Hundred project hydrodynamical simulations: Hydrodynamical weak-lensing cluster mass biases and richnesses using different hydro models.}   
   \subtitle{}

   \author{C. Giocoli
     \inst{1,2}\fnmsep\thanks{\email{carlo.giocoli@inaf.it}},
     G. Despali\inst{3,1,2},
     M. Meneghetti\inst{1,2},
     E. Rasia\inst{4,5,6},
     L. Moscardini\inst{3,1,2},
     S. Borgani\inst{7,4,5,8,9}
     G. F. Lesci\inst{3,1},
     F. Marulli\inst{3,1,2},
     W. Cui\inst{10,11,12},
     G. Yepes\inst{10,11}
   }

   \institute{
     INAF-Osservatorio di Astrofisica e Scienza dello Spazio di Bologna, Via Piero Gobetti 93/3, 40129 Bologna, Italy
     \and
     INFN -- Sezione di Bologna, viale Berti Pichat 6/2, I-40127 Bologna, Italy
     \and
     Dipartimento di Fisica e Astronomia "Augusto Righi", Alma Mater Studiorum Università di Bologna, via Gobetti 93/2, I-40129 Bologna, Italy        
     \and
     INAF -- Osservatorio Astronomico di Trieste, via Tiepolo 11, I-34131, Trieste, Italy
     \and
     IFPU, Institute for Fundamental Physics of the Universe, Via Beirut 2, 34014 Trieste, Italy
     \and 
     Department of Physics; University of Michigan, Ann Arbor, MI 48109, USA
     \and 
     Dipartimento di Fisica, Universit\`a di Trieste, Sez. di Astronomia, via Tiepolo 11, I-34131 Trieste, Italy  
     \and 
     ICSC - Italian Research Center on High Performance Computing, Big Data and Quantum Computing, via Magnanelli 2, 40033, Casalecchio di Reno, Italy   
     \and 
     INFN -- Instituto Nazionale di Fisica Nucleare, Via Valerio 2, I-34127, Trieste, Italy
     \and
     Departamento de F\'{i}sica Te\'{o}rica, M\'{o}dulo 15, Facultad de Ciencias, Universidad Aut\'{o}noma de Madrid, E-28049, Madrid, Spain 
     \and 
     Centro de Investigaci\'{o}n Avanzada en F\'{i}sica Fundamental (CIAFF), Facultad de Ciencias, Universidad Aut\'{o}noma de Madrid, E-28049, Madrid, Spain 
     \and
     Institute for Astronomy, University of Edinburgh, Royal Observatory, Edinburgh EH9 3HJ, UK
   }

\date{Received {\color{violet}{\it \underline{Space: 1999}}}; Accepted {\color{violet}{\it \underline{2001: A Space Odyssey}}}}
   
   \abstract
       {The mass of galaxy clusters estimated from weak-lensing observations is affected by projection effects, leading to a systematic underestimation compared to the true cluster mass. This bias varies with both mass and redshift. Additionally, the magnitude of this bias depends on the criteria used to select clusters and the spatial scale over which their mass is measured.

         In this work, we leverage state-of-the-art hydrodynamical simulations of galaxy clusters carried out with        \texttt{GadgetX} and \texttt{GIZMO-SIMBA} as part of the          Three Hundred project. We used them to quantify weak-lensing
         mass biases with respect also to the results from dark matter-only simulations. We also investigate how the biases of the weak-lensing mass estimates propagate into the richness-mass relation.}
   {We aim to shed light on the effect of the presence of baryons on the weak-lensing mass bias and also whether this bias depends on the galaxy formation recipe; in addition, we seek to model the richness-mass relation that can be used as guidelines for observational experiments for cluster cosmology.}
   {We produced weak-lensing simulations of random projections to model the expected excess surface mass density profile of clusters up to redshift $z=1$. We then estimated the observed richness by counting the number of galaxies in a cylinder with a radius equal to the cluster radius and correcting by large-scale projected contaminants. 
   We adopted a Bayesian analysis to infer the weak lensing cluster mass and concentration.}
   {We derived the weak-lensing mass-richness relation and found consistency within 1$\sigma$ uncertainties across hydrodynamical simulations. The intercept parameter of the relation is independent of redshift but varies with the minimum of the stellar mass used to define the richness value. At the same time, the slope is described by a second-order polynomial in redshift, which is relatively constant up to $z = 0.55$. The scatter in observed richness at a fixed weak-lensing mass, or vice versa, increases linearly with redshift at a fixed stellar mass cut. As expected,  we observed that the scatter in richness at a given true mass is smaller than at a given weak-lensing mass.  Our results for the weak-lensing mass-richness relation align well with SDSS redMaPPer cluster analyses when adopting a stellar mass cut of \( M_{\rm star,\,min} = 10^{10} \, h^{-1}  M_\odot \). Finally, we present regression parameters for the true mass–observed richness relation and highlight their dependence on redshift and stellar mass cut, offering a framework for improving mass–observable relations essential for precision
     cluster cosmology.}
   {}
   
   \keywords{dark matter -- weak gravitational lensing -- clusters of
     galaxies -- hydrodynamical simulations }
  \titlerunning{Hydrodinamical WL mass biases}
  \authorrunning{Giocoli et al. 2024}
  \maketitle

\section{Introduction}

Different wide-field observational facilities
\citep{planck16a,descosmo22,desicosmo24} and dedicated observations of
galaxies and clusters \citep{bergamini23,diego24} have highlighted
that dark matter (DM) is the major matter component in our
Universe. It is distributed along filamentary structures, walls, and
nodes where clusters of galaxies are embedded
\citep{malavasi17,libeskind18,santiago-bautista20,feldbrugge24,hoosian24,zhang24}.

Following the standard scenario of structure formation, cosmic
structures form hierarchically, and their average mass assembly
proceeds monotonically over time
\citep{tormen98a,tormen98b,vandenbosch02,wechsler02,giocoli07}. Galaxy
clusters are the last virialised structures to form, and DM dominates
their mass content, while their central region is affected by different non-linear dynamic
processes \citep{springel01b,tormen04} triggered by the baryonic
physics \citep{cui16,jake17}.

Baryons constitute the visible component of the clusters and can be
observed in different wavelengths using facilities from the ground and
space. In X-ray, clusters appear smooth since the X-ray photons are
emitted via the Bremsstrahlung thermal emission of the diffuse and hot
plasma, called intracluster medium. In the visible and near-infrared
bands, they instead appear as a clumpy distribution of knots for the
integrated emissions from individual stars in their hosting galaxies.

Due to their privileged role in the hierarchical structure formation
scenario and their relatively easy detectability, clusters are an
important cosmological probe
\citep{bahcall97,costanzi14,planckxx}. Indeed, their mass distribution
as a function of redshift, called the mass function, depends on
several cosmological parameters that can be constrained once the observed function is compared with theoretical expectations derived
from numerical simulations \citep{tinker08,despali16,castro23}.

 Clusters act as powerful gravitational lenses, bending light from background sources \citep{meneghetti08,meneghetti10b,natarajan24}. In strong lensing near the cluster centre, sources are highly magnified, distorted, and multiply imaged. In weak lensing, farther from the centre, distortions are subtle, requiring a statistical analysis of many galaxies to estimate the cluster mass. Since lensing is independent of the cluster’s dynamical state, weak lensing is widely used in cosmological surveys 
\citep{abbott20,ansarinejad24,bocquet24a,bocquet24b,giocoli24b,grandis24,sereno24,ingoglia24,salcedo24,kleinebreil25}.
to provide unbiased estimates of the projected matter distribution  \citep{bartelmann01,bartelmann10}.   
Nevertheless, mostly due to projection effects, the recovered lensing
mass has, on average, a low bias with respect to the actual
three-dimensional one
\citep{meneghetti08,becker11,giocoli12c,giocoli14,giocoli24}; however, it is worth mentioning that for an optically selected sample, projection effects may tend to 
boost the recovered weak-lensing mass \citep{abbott20,wu22}. Many
dedicated works have been carried out based on state-of-the-art
hydrodynamical simulations to assess the reliability of the recovered
weak-lensing (hereafter also WL) mass \citep{grandis19,grandis21}.
However, one aspect that has received little attention is the precise quantification of how baryonic physics influences the weak-lensing reconstruction of cluster mass as a function of redshift. Since weak lensing is a key tool for estimating cluster masses in cosmological surveys, understanding these effects is essential for minimising biases in mass measurements. The collisional nature of baryonic matter, combined with processes such as radiative cooling, star formation, and energy feedback from stars and AGN, alters the distribution of baryons and, in turn, affects the total mass distribution (Rasia et al. in prep.). For example, while cooling leads to a contraction of haloes through adiabatic processes \citep{gnedin04}, (active galactic nuclei) AGNs feedback can instead cause a slight expansion. These competing effects modify the overall matter distribution within haloes, influence cluster concentrations, and ultimately impact the lensing signal. If not properly accounted for, these modifications can introduce systematic biases in weak-lensing mass estimates, affecting the accuracy of cluster-based cosmological constraints. 
At the same time, different numerical implementations of star formation
in simulations can produce peaked subhaloes with a higher and
concentrated stellar distribution \citep{meneghetti23, li23}. In this
way, subhaloes become more resistant to disruption within the cluster
environment, with respect to a DM-only simulation
\citep[e.g.][]{dolag09}, thus increasing the normalisation of the
subhalo mass
function \citep{despali17b,ragagnin22,srivastava24}. Clearly, the
amount, scale-dependence, and timescale over which such processes
alter the total matter distribution in cluster-sized haloes and in
their substructures are determined by the details of the
sub-resolution processes that ultimately determine galaxy formation
in simulations.
The investigation of the weak lensing mass biases is essential for the exploitation of future wide-field surveys, which are aimed at accurate weak-lensing reconstruction of cluster masses to derive robust cosmological posteriors from
cluster number counts. This is the first issue that we tackle in
this paper. We compare the biases derived from three sets of
simulations from the Three Hundred collaboration: one carried out with
DM particles only, and the other two performed with different
hydrodynamical codes and sub-grid modules.  We emphasize that it is beyond the scope of this study to quantify the effects due
to shear bias, photometric redshift errors, and cluster
mis-centring. On the other hand, we include the effect of
contamination from nearby structures included within $\pm 5$ Mpc along
the line of sight with respect to the cluster centre. This scale represents the optimal compromise to avoid including low-resolution particles considering the size of the resimulated cluster region \citep{cui18}.
The mass of individual systems derived from gravitational lensing is,
as said, projected, and its three-dimensional reconstruction requires
both high-quality data and, eventually, calibration using
state-of-the-art dedicated numerical simulations
\citep{ragagnin24}. 

Direct mass estimation is limited to high signal-to-noise objects and cannot be applied to the vast data sets from upcoming missions. A solution to this is to use mass proxies tightly correlated with total mass. Multi-band observations help establish well-calibrated mass-observable relations, thus enabling mass estimates for large cluster samples. A key proxy is cluster richness—measured by counting satellite galaxies above a brightness or mass threshold \citep{ivezic08,ivezic09}.

This method has been validated, for example, by \citet{costanzi21}, who calibrated the richness-mass relation using (the Dark Energy Survey) DES and (the South Pole Telescope) SPT data, and it has shown consistency with other cosmological probes. Accurate calibration depends on the photometric identification of galaxies and cluster selection functions \citep{sartoris16}. Prior studies 
\citep{lima05,sartoris10,carbone12}
often assumed a fixed precision rather than directly deriving scaling parameters. \citet{andreon12} demonstrated that accounting for selection effects improves precision, while \citet{andreon16} compiled a cluster mass catalogue with 0.16 dex accuracy, surpassing X-ray or SZ methods. This data set improves consistency across studies by refining cluster centre, redshift, and contamination measurements.
In a more recent analysis, \citet{lesci22,lesci22b}
analysed cluster counts in the AMICO KiDS-DR3 catalogue jointly
constraining cosmological parameters and the cluster mass-observable
scaling relation using intrinsic richness as the observable linked to
cluster mass. The sample includes thousands of clusters with a large richness up to $z =0.6$. Following \citet{bellagamba19}, for the
weak-lensing analysis, \citet{lesci22,lesci22b} corrected for incompleteness and
impurities using a mock catalogue. Their model for cluster counts
accounts for redshift uncertainties and incorporates the intrinsic
scatter in the scaling relation, combining the likelihood functions to
constrain both cluster counts and parameters, and it determines the mass-observable scaling relation.

Using the two hydro-simulation sets of The Three Hundred, the second
goal of this work is to investigate the scaling relation between
richness and mass obtained from WL data. In particular, we focus
on how the relation changes according to the stellar mass cut used to
select the galaxy members and on the observational biases due to the
weak-lensing mass calibration.  Specifically, by using a mass-selected
sample, we aim to set priors on the parameters defining the
richness-mass relation, employing cluster richness as our mass proxy. 

The structure of this paper is as follows.  In Section \ref{sec_1}, we
describe the simulations employed in this study and the methodology
used to construct the weak-lensing signal from simulated
clusters. Additionally, we present our findings on weak-lensing mass
biases, examining their dependence on the hydrodynamical codes and
their redshift evolution.  In Section \ref{sec_2}, we discuss the
weak-lensing mass-observed richness relations -- and its inverse,
summarising them into a final model and comparing our findings with
existing results in the literature. In Section \ref{sec_3}, we
calibrate the scaling relation between true mass and observed
richness. Finally, we summarise our main conclusions in Section
\ref{sec_sc}.

All logarithms in this work are on base ten unless otherwise
indicated.

\section{Weak lensing: The simulations and the model}
\label{sec_1}
\begin{figure*}
  \centering
  \includegraphics[width=\hsize]{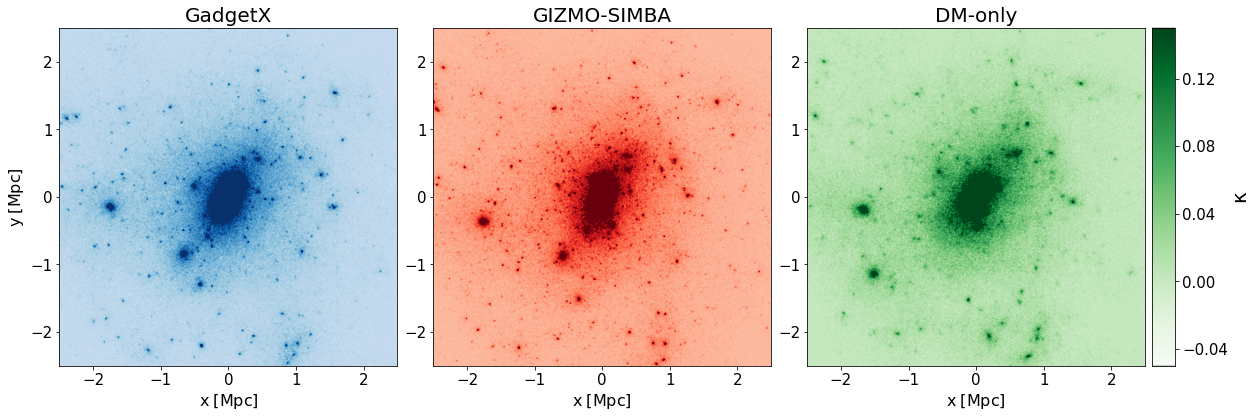}
  \caption{Simulated convergence maps of a cluster lens, namely with
    ID=4, at $z_l=0.22$ and considering $z_s=3$, obtained by collapsing
    the total mass along the z-projection. The left and the central
    panels show the results using two different hydro-solvers
    \texttt{GadgetX} and \texttt{GIZMO-SIMBA}, respectively. In the
    right panel, we display the convergence map of the same cluster
    projection simulated using only collisionless dark matter
    particles, \texttt{DM-only} run. The assumed field of view is 5
    Mpc by side, collapsing $\pm$ 5 Mpc matter along the
    line-of-sight.\label{figMaps} }
\end{figure*}

\subsection{The Three Hundred project runs}

We build up our weak-lensing cluster images using simulated regions by
the Three Hundred Collaboration \citep{cui18,cui22}.  A total sample
of 324 clusters has been mass-selected ($M_{200}>8\times 10^{14}\,
h^{-1}M_{\odot}$) at $z=0$ in the MultiDark MDPL2 cosmological N-Body
simulation \citep{klypin16}. The original and resimulated runs have
been performed assuming the cosmological parameters from the
\citet{planck16a} $\Omega_{\rm m}=0.307$, $\Omega_{\rm b}=0.048$,
$\Omega_{\rm \Lambda}=0.693$, $h=0.678$, $\sigma_8=0.823$ and $n_{\rm
  s}=0.96$, following the evolution of $3840^3$ collisionless
particles with mass $m_{\rm DM}=1.5 \times 10^9\,h^{-1}M_{\odot}$ on a
box with 1 $h^{-1}$ Gpc on a side.  For each of the 324 regions --
defined as being five times the virial radius of the cluster position
on their centre, initial conditions with multiple levels of mass
refinements were generated using the \texttt{GINNUNGAGAP}
code\footnote{\url{https://github.com/ginnungagapgroup/ginnungagap}},
and dividing the mass content in baryonic and DM mass
according to the adopted cosmological baryon fraction, leading to:
$m_{\rm gas}=12.7\times 10^8\,h^{-1}\,M_{\odot}$ and $m_{\rm
  dm}=2.36\times 10^8\,h^{-1}\,M_{\odot}$, respectively.  Starting
from redshift $z=120$, when the initial conditions have been
generated, each zoom-in simulation has been run with two different
hydro-solver algorithms: \texttt{GadgetX} \citep{rasia15} and
\texttt{GIZMO-SIMBA} \citep{dave19,cui22}.  \texttt{GadgetX}
simulations have been performed with a ‘modern’ smoothed-particle
hydrodynamics (SPH) code based on \texttt{Gadget-3} that includes
artificial thermal diffusion, time-dependent artificial viscosity,
high-order Wendland C4 interpolating kernel, and wake-up scheme 
\citep{beck16a}.  The GIZMO-SIMBA runs are completed with the GIZMO
code \citep{hopkins15} with the state-of-the-art galaxy formation
subgrid models following the SIMBA simulation \citep{dave19}. The
meshless finite mass (MFM) solver implemented in GIZMO evolves gas
particles with a precise treatment of shocks and shear flows,
eliminating the need for artificial viscosity.  In addition, starting
from the same initial conditions and setting exactly the same initial
sampling of phase space, a collisionless \texttt{DM-only} run has been
produced using \texttt{Gadget-3}. This run is used as our
reference.

These simulated regions have been previously used to investigate
various properties of both the DM and the baryon content. Relevant to
this paper, we mention the studies on the influence of the environment
on galaxy properties \citep{wang18}, the cluster splashback radius
characterization \citep{rogan20,knebe20}, the baryon profile
\citep{li23} and the hydrodynamical mass bias \citep{gianfagna23}, the
strong lensing properties of the clusters \citep{vega-ferrero21} and
their satellite galaxies \citep{meneghetti23,srivastava24}, and the
weak-lensing mass bias \citep{giocoli24}.

At all the analysed snapshots, nine from $z=0.12$ to $z=0.94$, the
main properties of each cluster are computed via the Amiga Halo Finder
\citep{knollmann09} and include the total mass $M_{200}$, defined as
the mass within the radius $R_{200}$ which encloses 200 times the
critical density of the universe $\rho_{\rm c}(z)$ at the
corresponding redshift,
\begin{equation}
M_{200} = \dfrac{4 \pi}{3} R_{200}^3\, 200 \; \rho_{\rm c}(z)\,.
\end{equation}
In particular, we define $M_{200,\rm DM}$ to be the cluster mass in
the \texttt{DM-only} run.

As mentioned, to model the lens, we also need the cluster
concentration, $c_{200}$ that adopting a Navarro-Frenk-White
(hereafter NFW) density profile \citep{navarro96,navarro97} is defined
as the ratio between $R_{200}$ and the scale radius $r_{\rm s}$,
defined as the radius at which the logarithmic derivative of the
density profile is equal to $\mathrm{d}\log \rho /\mathrm{d}\log r =
-2 $.  Following the NFW density profile parametrisation,
\citet{prada12} describe the halo concentration in terms of the
velocity ratio as follows:
\begin{equation}
  \dfrac{V_{\rm max}}{V_{200}} = \sqrt{\dfrac{0.216 \; c_{200}}
    {\ln(1+c_{200}) -c_{200}/(1+c_{200})} }\,,
\label{eqPrada}
\end{equation}
where $V_{200} = \sqrt{\dfrac{G M_{200}}{R_{200}}}$ represents the
halo virial circular velocity and $V_{\rm max}$ the maximum circular
velocity. We derive $c_{200}$ by solving numerically
Eq.~\ref{eqPrada}.
    
\subsection{Cluster weak-lensing simulations}

To create and analyse the simulated lensing images, we follow the same procedure of \citet{giocoli24}, but here we extend the sample to both
the \texttt{GIZMO-SIMBA} and the \texttt{DM-only} versions.
For each of the 324 central clusters of the resimulated regions, we
build three simulated excess surface mass density profiles, collapsing
each time the particles into a single lens plane along the simulated
box axes; this guarantees that the three projections are random with respect to the three-dimensional shape of the clusters. We repeat the
same procedure for all nine different snapshots, consistently
accounting for the redshift evolution of expected background sources
beyond the clusters.  For each projection along the simulation axes,
we select particles in a slice of depth $\pm 5\, \mathrm{Mpc}$
($3.4\,h^{-1}\,\mathrm{Mpc}$) in front and behind the cluster and $\pm
2.5\, \mathrm{Mpc}$ ($1.7\,h^{-1}\,\mathrm{Mpc}$) from the projected
cluster centre in the considered plane of the sky. The depth of the
projection along the line of sight has been chosen as a compromise to
include the neighbouring correlated structure and avoid including
low-resolution particles in the lensing simulated maps. Our method
used to smooth the mass distribution considers a smoothing length
equal to the distance of the $n^{\rm th}$ neighbour particle, with
$n=80$. This is done for each particle species weighted according to
their masses -- dark matter, gas, star, and black hole -- using
\texttt{Py-SPHViewer} \citep[for more details, we refer
  to][]{benitez15} on a grid with a pixel resolution of $2048 \times
2048$ and 5 Mpc on a side.
  
The convergence, $\kappa$, is obtained from the mass map by dividing
the mass per pixel by its associated area to obtain the surface
density, $\Sigma(\vec{\theta})$, and by the critical surface density,
$\Sigma_{\rm crit}$ \citep{bartelmann01} that can be read as
\begin{equation}
  \Sigma_{\rm crit} \equiv \dfrac{c^2}{4 \pi G} \dfrac{D_{\rm s}}{D_{\rm l}
    D_{\rm ls}}\,,
  \label{eq_scrit}
\end{equation}
where $D_{\rm l}$, $D_{\rm s}$, and $D_{\rm ls}$ are the
observer-lens, observer-source and source-lens angular diameter
distances, respectively; $c$ represents the speed of light and $G$ the
universal gravitational constant. Our reference weak-lensing maps have
been created assuming a fixed source redshift $z_{\rm s}=3$.

As an example, in Fig.\ref{figMaps}, we show the convergence map
obtained by collapsing the particles along the $z$-direction of the
re-simulated box, considering one cluster at redshift $z_l=0.22$: from
left to right, we show the convergence obtained from the
\texttt{GadgetX}, \texttt{GIZMO-SIMBA}, and \texttt{DM-only} run. Note
that the colour code shown in this figure (blue, red, and green
respectively for \texttt{GadgetX}, \texttt{GIZMO-SIMBA}, and
\texttt{DM-only}) are used throughout the paper.  In all cases,
the large-scale projected matter density distribution looks relatively
similar, while the position of substructures differs due to the
slightly different timings of evolution introduced by the presence of
the collisional component.

From the convergence $\kappa$, we can write the lensing potential
$\psi$, from the two-dimensional Poisson equation, as:
\begin{equation}
  \Delta_{\theta} \psi(\vec{\theta}) = 2 \, \kappa
  (\vec{\theta})\,,
  \label{eqpsi}
\end{equation}
that we numerically solve in Fourier space using the Fast Fourier
Transform (FFT) method.  From Eq.~\ref{eqpsi}, we can detail the two
components $(\gamma_1,\gamma_2)$ of the pseudo-vector shear, as:
\begin{eqnarray}
  \gamma_1(\vec{\theta}) &=& \dfrac{1}{2} \left(\dfrac{\partial^2
    \psi(\vec{\theta})}{\partial x^2} - \dfrac{\partial^2
    \psi(\vec{\theta})}{\partial y^2}
  \right)\;,\\ \gamma_2(\vec{\theta}) &=& \dfrac{\partial^2
    \psi(\vec{\theta})}{\partial x \,\partial y} \,,
\end{eqnarray}
where $x$ and $y$ represent the two components of the vector
$\vec{\theta}$.

We use these equations to build shear maps that are then analysed in
the next sections.  For each cluster, we use the shear maps to
simulate an observed surface mass density profile, $\Delta \Sigma$, by
randomly sampling the field of view with a given number density of
expected background sources. Notice that the real observable is the
reduced shear, $g={\gamma}/({1-\kappa})$, which, in the weak lensing
regime, we can approximate as $g \simeq \gamma$.  We assume a
background density of sources for weak lensing normalised to the total
value of 30 galaxies per square arcmin, with a peak around $z=1$
\citep{boldrin12,boldrin16,giocoli24}.
The excess surface mass density, $\Delta \Sigma$, shown in
Fig.\ref{figDeltaSigma} is defined as
$\Delta \Sigma (\theta) = \bar{\Sigma}(<\theta) - \Sigma(\theta)
\equiv \Sigma_{\rm crit} \gamma_{\rm t}(\theta)$, with
\begin{equation}
\gamma_{\rm t} (\theta_i) =- \gamma_1(x_i,y_i) \cos(2 \phi_i) -
\gamma_2(x_i,y_i) \sin(2 \phi_i)\,,
\end{equation}
where $(0,0)$ is the centre of the cluster by construction, $\theta_i
= \smash{(x_i^2+y_i^2)^{1/2}}$ and $\phi_i = \arctan \left( y_i/x_i
\right)$.

The associated error bars account for the intrinsic shape of the
background galaxies and the uncertainty on the mean value within the
annulus (see \citet{giocoli24}), which can be read as:
\begin{equation}
\sigma_{\Delta \Sigma} = \sqrt{ \sigma_{ \langle \Delta \Sigma \rangle
  }^2 + \Sigma_{\rm crit}^2 \dfrac{\sigma_{\rm e}^2 }{n_{\rm g} \; \pi
    \left( \theta_2^2 - \theta_1^2 \right) } }\,,
\end{equation}
where $\sigma_{\rm e} = 0.3$
\citep{hoekstra04,hoekstra11,kilbinger14,blanchard20} is the
dispersion of the shape of background source galaxies, and $\theta_1$
and $\theta_2$ are the lower and upper bounds of the considered radial
annulus.  From Fig. \ref{figDeltaSigma}, we observe that for this
specific cluster, the run performed with the \texttt{GadgetX} code has
a higher excess surface mass density than the other runs (GIZMO-SIMBA
and DM-only) as evident from the bottom sub-panel of the figure.

\subsection{Profile model} \label{subsec:2.3}

In order to model the simulated weak-lensing profile, we adopt a
smoothly truncated NFW density profile \citep[BMO,][]{baltz09},
defined as:
\begin{equation}
  \rho(r_{\rm 3D}|M_{200},c_{200},R_{\rm t})=\rho_{\rm NFW}(r_{\rm
    3D}|M_{200},c_{200}) \left(\frac{R_{\rm t}^2}{r_{\rm 3D}^2+R_{\rm
      t}^2}\right)^2 \,,\label{trunc.NFW}
\end{equation}
with $R_{\rm t} = t\,R_{200}$ with $t$ defined as the truncation
factor. Following the results by \cite{oguri11b}, \cite{bellagamba19},
and \cite{giocoli21}, we adopt a truncation radius $R_{\rm
  t}=3\,R_{200}$.  The total mass enclosed within $R_{200}$,
$M_{200}$, can be thought of as the normalisation of the model and as
a mass proxy of the true enclosed mass of the DM halo hosting
the cluster \citep{giocoli12a}. Writing $r_{\rm 3D}^2$ as the sum in
quadrature between the sky projected coordinate $r=D_{\rm d} \theta$
and the line-of-sight $\zeta$ coordinate, and integrating along
$\zeta$, we can write
\begin{equation}
\Sigma(r | M_{200},c_{200},R_{\rm t}) = \int_0^{\infty} \rho(r,\zeta |
M_{200},c_{200},R_{\rm t})\, \mathrm{d} \zeta\,.
\label{ptrunc.NFW}
\end{equation}
The differential excess surface mass density can then be written as:
\begin{equation}
\Delta \Sigma (r) = \dfrac{2}{r^2} \int_0^{r} r' \; \Sigma (r') \,\mathrm{d} r'  - \Sigma(r)\,.
\end{equation}
In order to model the data, we performed a Bayesian analysis using
Monte Carlo Markov Chain approach, assuming a Gaussian log-likelihood
between the model and the data
\footnote{\url{https://federicomarulli.github.io/CosmoBolognaLib/Doc/html}}
\citep{marulli16} that can be read as:
\begin{equation}
\mathcal{L} \propto \exp \left( - \dfrac{1}{2} \chi^2 \right)\,,
\label{eqLikelihood}
\end{equation}
where 
\begin{equation}
\chi^2 = \sum_{i} \left( \dfrac{\Delta \Sigma_{i}(r_i) - \Delta
  \Sigma_{\rm model}(r_i) }{\sigma_{\Delta \Sigma_i}} \right)^2\,,
  \end{equation}
and the sum extends to the number of radial bins.  We set uniform
priors for $\log (M_{200}/ [h^{-1}\,M_{\odot}]])\in[12.5,\,16]$ and
$c_{200}\in[1,\;15]$, and let each MCMC chain run for $16\,000$
steps.  

In our simulated weak-lensing analyses, we assume a diagonal covariance matrix, neglecting off-diagonal contributions that arise in real observations. This assumption is motivated by the fact that, in a controlled simulation setting, we primarily account for shape noise as the dominant source of uncertainty. Off-diagonal terms in the covariance matrix are typically introduced by correlated large-scale structure (LSS) along the line of sight \citep{hoekstra13,gruen15}, sample variance \citep{singh17}, and survey-specific systematics \citep{maccrann22}. These effects contribute significantly to the uncertainty in cluster mass estimates but are absent in our idealised approach, which isolates weak-lensing signal uncertainties from observational complexities.
While real weak-lensing analyses incorporate these additional sources of covariance, leading to significant off-diagonal terms \citep{schneider02,takada07,schneider22}, our simulations focus on an idealised scenario where such contributions are not explicitly modelled. This allows for a clearer interpretation of the primary sources of statistical uncertainty while avoiding biases introduced by imperfect modelling of LSS or systematic effects. Nevertheless, it is important to acknowledge that omitting off-diagonal terms may lead to an underestimation of the total uncertainty in weak-lensing mass estimates 
\citep{hoekstra03}. Future work incorporating more realistic simulations, including correlated LSS structures and survey-dependent systematics, will be necessary to fully quantify these effects.  
 
\begin{figure}
  \centering
  \includegraphics[width=\hsize]{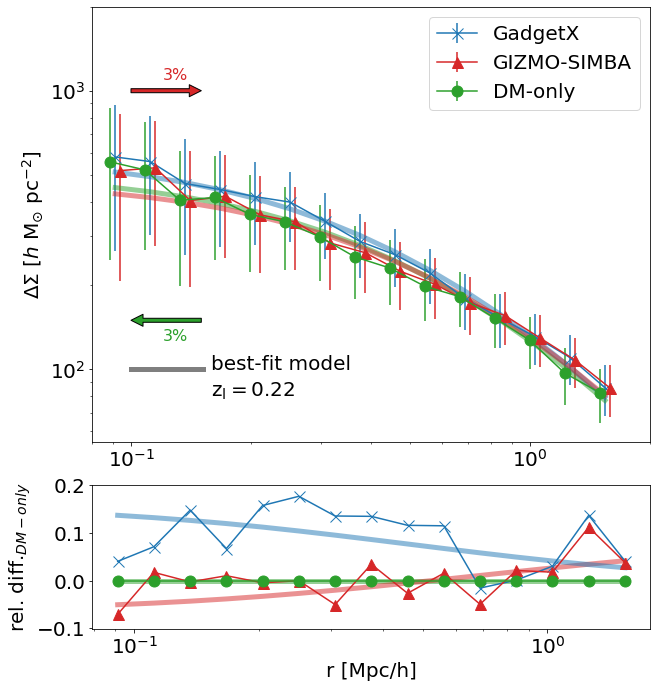}
  \caption{Excess surface mass density profile of the projections
    displayed in Fig.~\ref{figMaps}. The colours used for the data
    points, blue, red and green, correspond to the \texttt{GadgetX},
    \texttt{GIZMO-SIMBA} and \texttt{DM-only} run, respectively. The
    bottom panel displays the relative difference with respect to the
    \texttt{DM-only} case.  The best-fit models are shown using solid
    curves, coloured according to each considered
    case.\label{figDeltaSigma}}
\end{figure}

In Fig.~\ref{figDeltaSigma}, the solid curves show the best-fit result
as the model computed from the median values of the posterior
distributions for the same considered cluster as in the previous
figure.  The bottom sub-panel displays the relative difference between
the excess surface mass density measured --- and modelled -- in the
hydro runs versus the \texttt{DM-only} case.  From the figure, we can
observe that the \texttt{GadgetX} simulated clusters present a
steepening towards the centre, $10\%$ larger than the other two
cases. In contrast, the hydro run \texttt{GIZMO-SIMBA} profile is
quite close to the \texttt{DM-only} one.

\begin{figure}
  \centering
  \includegraphics[width=\hsize]{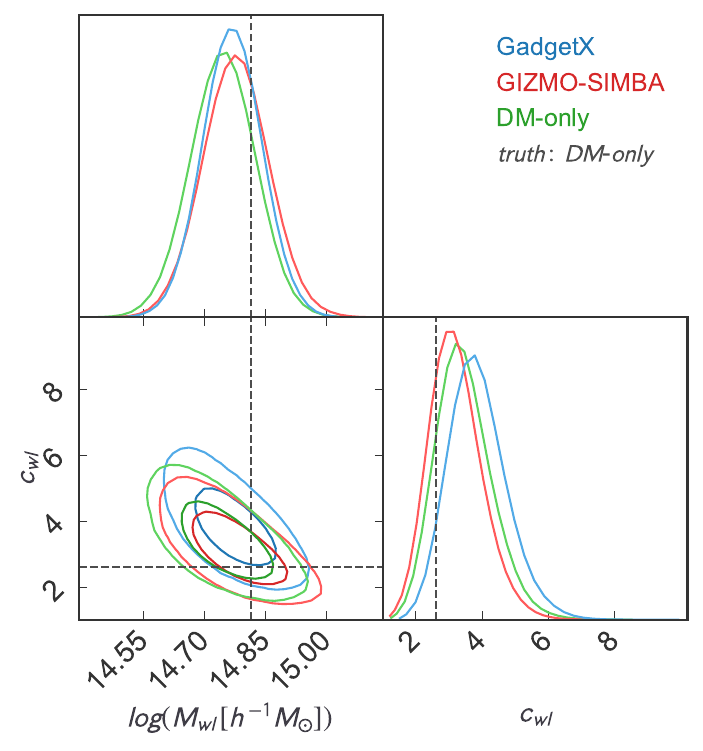}
  \caption{Posterior distributions of derived weak-lensing mass and
    concentration from the three weak-lensing simulations of the
    cluster of Fig.~\ref{figMaps} and Fig.~\ref{figDeltaSigma}. The
    dashed lines indicate the three-dimensional mass and concentration
    for the \texttt{DM-only} cluster run.  \label{figPosteriors}}
\end{figure}

In Fig~\ref{figPosteriors}, we show the posterior distributions of the
derived mass and concentration. We compare our results with respect to
the parameters of the \texttt{DM-only} cluster. In particular, while
all derived masses are consistent within $1\,\sigma$ uncertainties
with respect to the run with only dark matter (dashed vertical line), the concentration of
the \texttt{GadgetX} run turns out to be more than $1\,\sigma$ away
from the \texttt{DM-only} case (dashed horizontal line), as a consequence of the steeper
projected profile towards the centre, as displayed in
Fig.~\ref{figDeltaSigma}.

   
\begin{figure*}
  \centering
  \includegraphics[width=0.49\hsize]{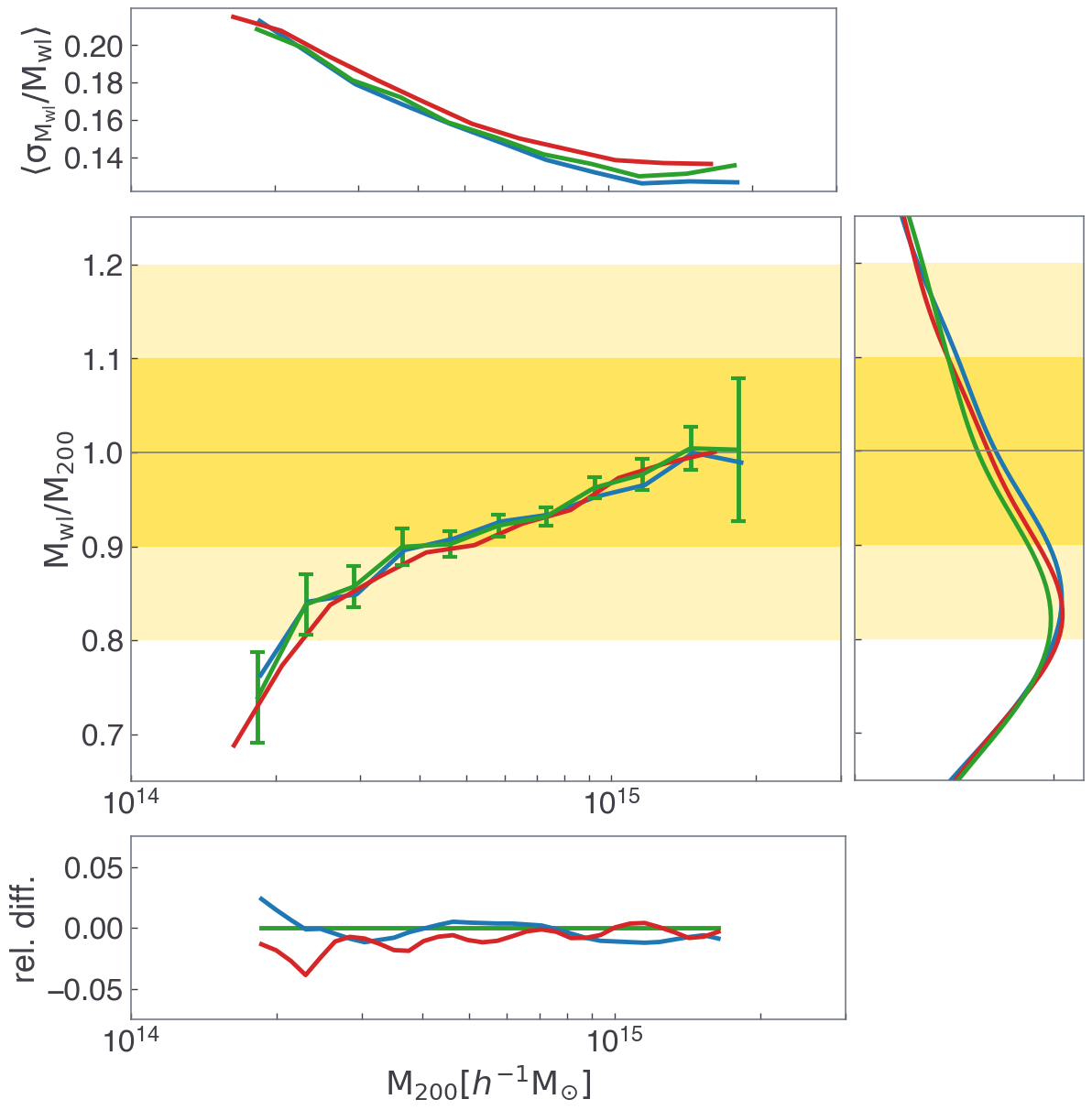}
  \includegraphics[width=0.49\hsize]{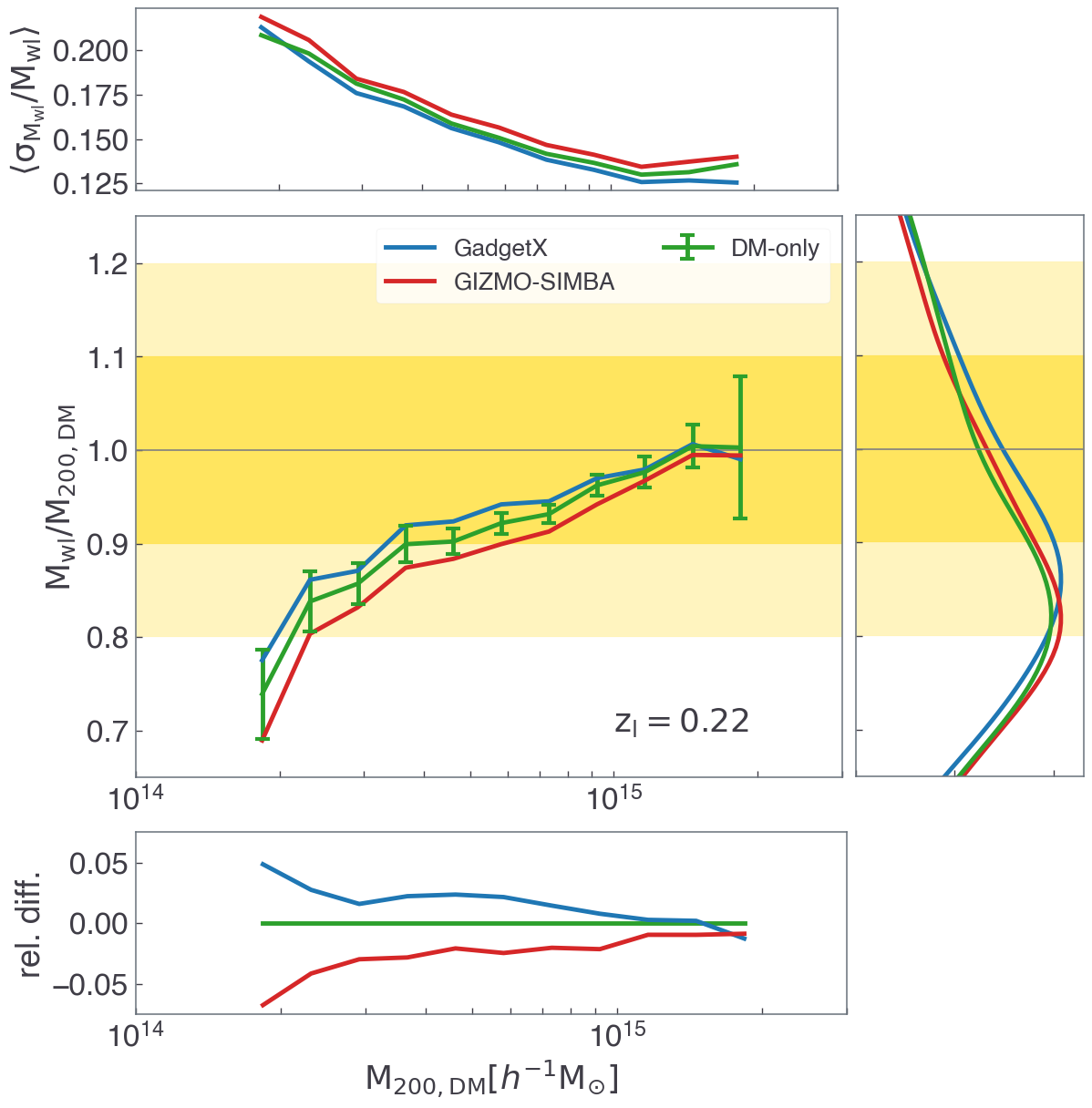}
  \caption{Average weak-lensing mass biases as a function of the
    corresponding three-dimensional masses. On the left, the weak
    lensing masses are relative to their respective total mass; on the
    right, they are always related to the total mass of the \texttt{DM-only} runs. The top and bottom sub-panels show the mean relative
    uncertainties and the relative differences with to the \texttt{DM-only} case as a function of considered three-dimensional reference mass, respectively. The
    right panels display the distribution of the weak-lensing mass
    biases over all cluster masses. The dark and light yellow bands show 
    the $10\%$ and $20\%$ differences, respectively. \label{figWLMbias}}
\end{figure*}

In Fig.~\ref{figWLMbias}, we now consider all $z=0.22$ maps and show
the average weak-lensing mass bias for all cluster projections with
respect to either their corresponding true mass (on the left) or the
true mass as measured in the \texttt{DM-only} simulation (on the
right), this latter typically used in the halo mass function
calibration \citep{sheth99b,tinker08,despali16}.  The right sub-panels
show the PDF of the data of the three sets of simulations, while the
top ones show the average relative uncertainties as a function of the
corresponding rescaled mass; the bottom panels present the relative differences
with respect to the \texttt{DM-only} mass biases. 
From the left figure, we see that despite the differences among the three simulation sets, they all provide the same answer for the weak-lensing bias and its scatter.
From the right figure, when we compare the hydro weak-lensing masses with the 
\texttt{DM-only} masses, we find that, on average, the two hydro-runs show a specular
trend, reflecting the results from the single cluster illustrated in the
previous figures. While the \texttt{GadgetX} weak-lensing masses
reduce the DM-only bias, the \texttt{GIZMO-SIMBA} ones tend to be
larger for smaller systems. Both codes have differences that tend to
vanish when $M_{\rm 200,\,DM} \gtrsim 10^{15}\,h^{-1}M_{\odot}$, as
displayed in the bottom sub-panel. This result and its dependence on
the mass is linked to the fact that even starting from the same
initial condition, \texttt{GIZMO-SIMBA} produces groups which are less
massive than \texttt{Gadget-X} since its strong feedback pushes out a
large fraction of the gas mass and thus slow down the overall matter
accretion.   
   
\begin{figure}
  \centering
  \includegraphics[width=\hsize]{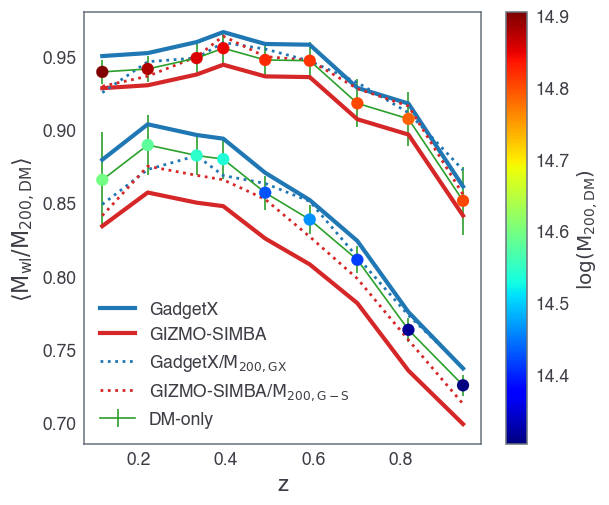}
  \caption{Average weak-lensing mass biases as a function of
    redshift. Data points with error bars, colour-coded according to
    their \texttt{DM-only} corresponding mass, show the dark
    matter-only run results. Blue and red lines refer to the
    \texttt{GadgetX} and \texttt{GIZMO-SIMBA} hydro simulation cases,
    respectively. {Dotted and solid lines show the bias in relation to
      the hydro mass and the DM-only mass,
      respectively.}\label{figWLMbiasZ}}
\end{figure}

\citet{giocoli24} have presented the average cluster weak-lensing mass
biases as a function of redshift, using the results of the
\texttt{GadgetX} simulations. To complement their findings, in
Fig. \ref{figWLMbiasZ}, we show the average weak-lensing mass biases
with respect to the dark matter-only simulation mass as a function of
redshift for the three simulation runs.  In order to underline the
dependence on the halo mass at each redshift, we split the cluster
sample into two subsamples, containing clusters with $M_{\rm
  200,\,DM}$ larger or smaller than $5\times
10^{14}\,h^{-1}M_{\odot}$. We colour-code the results by the
\texttt{DM-only} mass. In general, we note that more massive clusters
suffer from a smaller WL mass bias, as underlined by
\citet{giocoli24}, depending on the weak lensing signal-to-noise,
modulated by the considered source redshift distribution.  It is also
worth mentioning that the \texttt{GadgetX} and \texttt{GIZMO-SIMBA}
clusters have an opposite trend with respect to the \texttt{DM-only}
simulation: while the former ones are on average higher by $1.2\%$
($1.6\%$), the latter are ones lower by $1.2\%$ (3.7\%), for the more
massive (less massive) clusters with $M_{\rm 200,\,DM}\geq 5\times
10^{14}\,h^{-1}M_{\odot}$ ($M_{\rm 200,\,DM}< 5\times
10^{14}\,h^{-1}M_{\odot}$), respectively.
The dotted lines display the redshift evolution of the mass biases
with respect to the considered three-dimensional hydrodynamical, again
underlining that, in this case, the differences with respect to the corresponding true mass are only within a few per cent, between different simulations.

Before concluding this section, it is worth noticing that
\citet{lee18} have underlined that the biases at small masses could be
driven by the concentration priors and the anti-correlation correlation
between concentration and mass; moreover, while \citet{ragagnin24}
highlighted that this could be due to the poor NFW-fit for those
systems, \citet{giocoli14} and \citet{giocoli24} emphasised that the weak lensing signal-to-noise ratio and halo triaxiality could play the important role.

\section{Weak-lensing mass-richness relations}
\label{sec_2}

A key approach to improving mass estimates for large cluster samples is to use the weak-lensing mass–richness relation, where richness —- typically defined as the number of satellite galaxies in a cluster above a given luminosity or stellar mass threshold in a considered aperture -— acts as an indirect proxy for total mass. However, translating richness into an accurate mass estimate requires thorough calibration, as richness measurements are affected by selection effects, projection biases and photometric redshift uncertainties. Dedicated simulations are crucial for quantifying and correcting these biases to ensure accurate cosmological constraints from cluster counts.
In particular, richness estimates are affected by interloping galaxies falsely assigned to clusters due to projection along the line of sight. Simulations can model these effects, enabling corrections for contamination and biases in richness measurements. In addition, the weak-lensing mass-–richness relation may evolve with redshift, as well as the scatter,  due to changes in galaxy populations, merger activity, or dynamical relaxation timescales. Simulations allow for the testing of redshift-dependent scaling parameters to ensure accurate extrapolations in high-redshift cluster studies.

This section presents the weak-lensing mass-richness relation in our
simulated mass-selected cluster sample.
\begin{figure}
  \centering
  \includegraphics[width=\hsize]{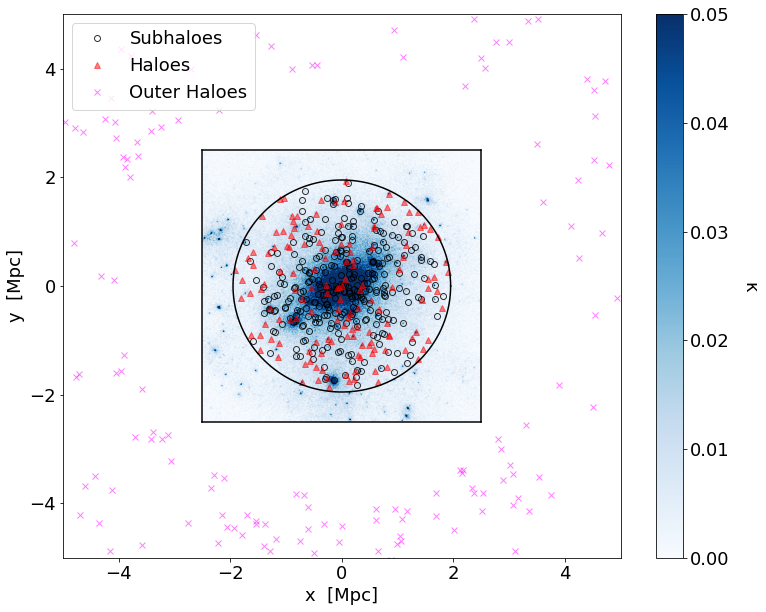}
  \caption{Halo and subhalo distribution in the projection $x$-$y$
    considering a field of view of 10 Mpc on a side and $\pm 5$ Mpc
    along the line of sight, $\Delta$H = 10 Mpc, for the
    \texttt{GadgetX} cluster run. Black circles show the location of
    the centre of all subhaloes within the halo radius $R_{200}$, red
    triangles are all haloes which, in projection, fall within the
    halo radius, and magenta crosses are all haloes that, in
    projection, lie in a ring between 2 and 3.5 $R_{200}$, which we
    term outer haloes. The blue map is the corresponding region where
    we simulate the weak-lensing cluster
    signal. \label{figSubsandhaloes}}
\end{figure}
\begin{figure*}
  \centering
  \includegraphics[width=0.49\hsize]{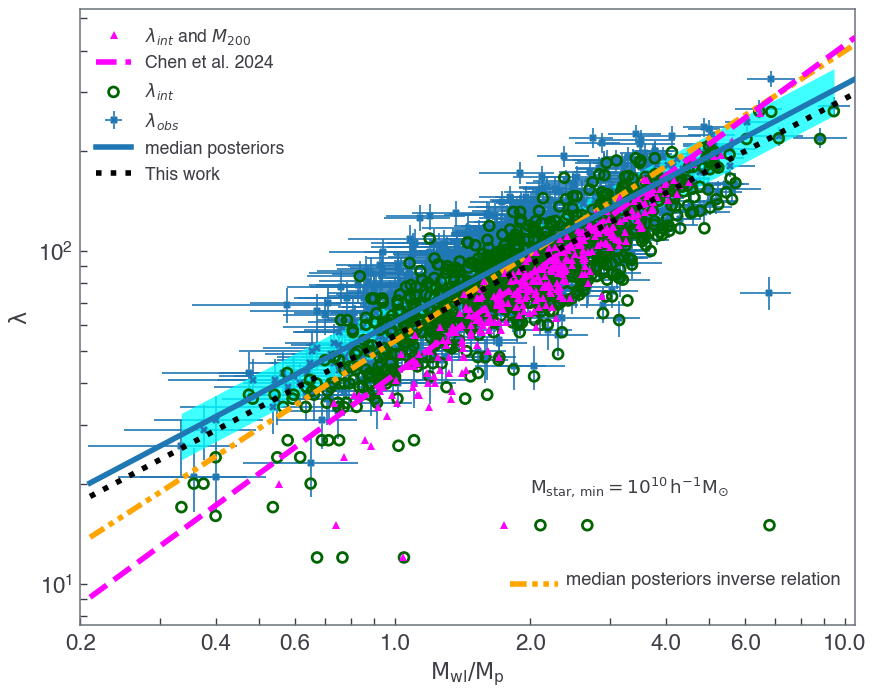}   
  \includegraphics[width=0.49\hsize]{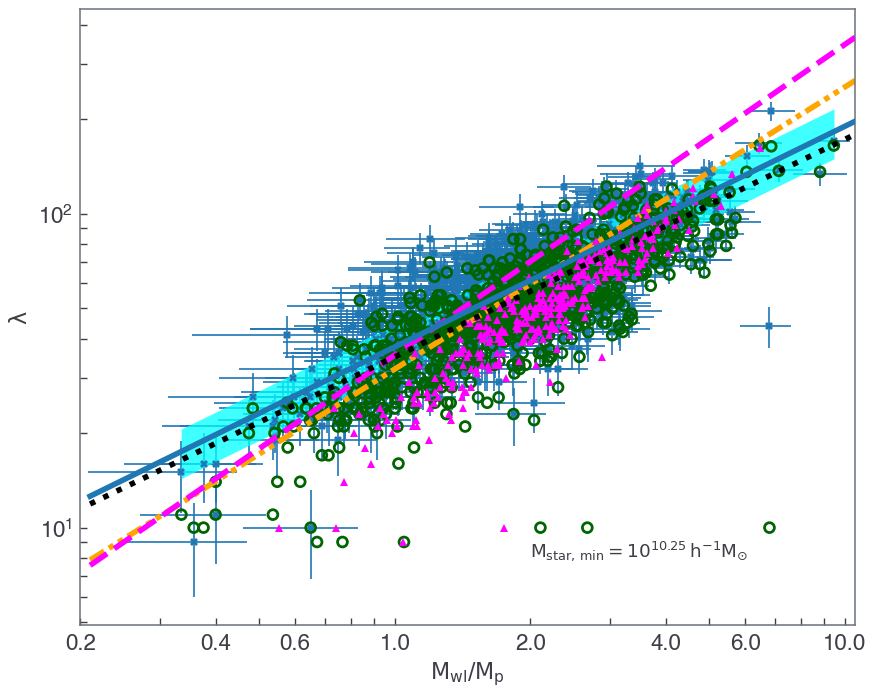}     
  \includegraphics[width=0.49\hsize]{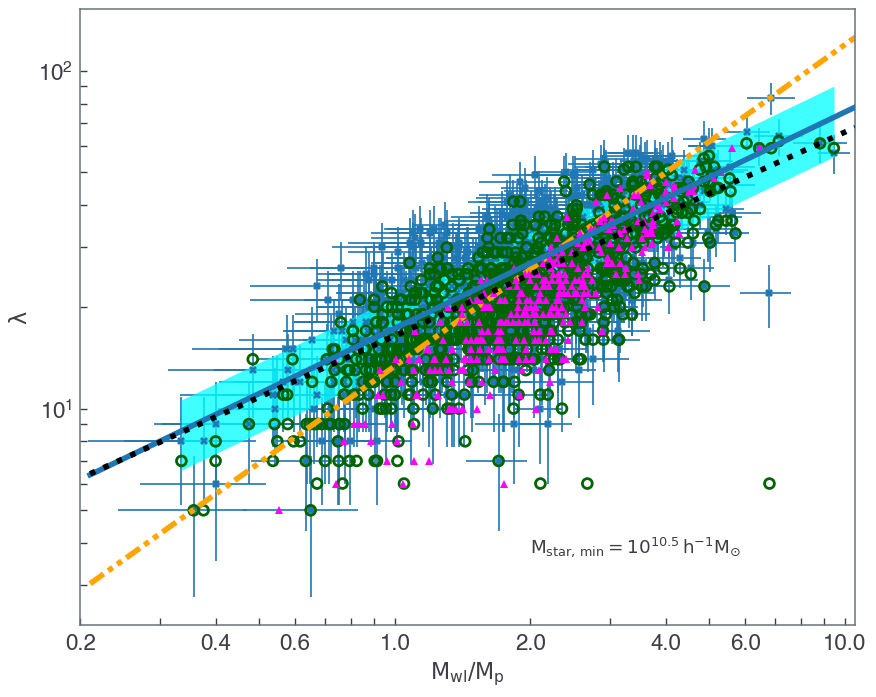}      
  \includegraphics[width=0.49\hsize]{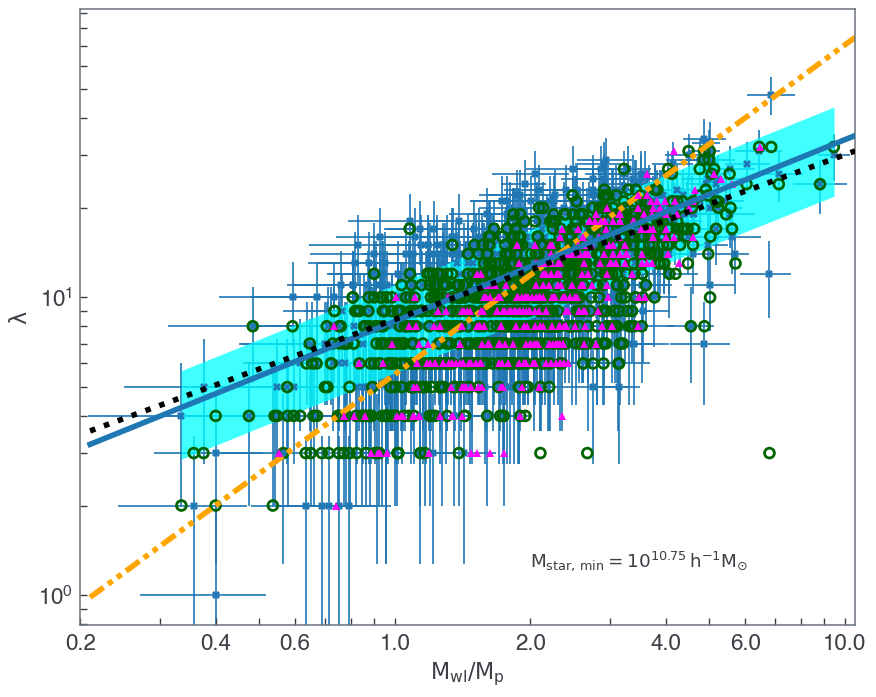}        
  \caption{Weak-lensing mass-observed richness relation measured for
    the clusters at redshift $z=0.22$ run with \texttt{GadgetX}.  The
    different panels consider different minimum stellar mass cuts
    $M_{\rm star, \,min}$, from $10^{10}$ to
    $10^{10.75}h^{-1}M_{\odot}$. The weak-lensing masses M$_{\rm wl}$
    are rescaled with respect to the pivot value $M_{\rm p} = 3\times
    10^{14}\,h^{-1}M_{\odot}$.  The blue points with the corresponding
    error bars show the results for the observed richness calculated
    as in Eq.\ref{eqrichness} and the derived weak-lensing mass from
    the tangential shear profile. For the magenta triangles, the mass
    is not M$_{\rm wl}$, but $M_{200}$; correspondingly we show the
    number $\lambda_{\rm int}$ of galaxies within $R_{200}$. The green
    circles show the results when using $M_{wl}$ and the number of
    subhaloes in a sphere with radius $R_{\rm 200}$. The solid blue line
    displays the results of our fit performed using MCMC on the blue
    data points, and in cyan, the $1\,\sigma$ uncertainty.  The orange
    dash-dotted line is the best fit of the inverse relation; see
    Eq.\ref{eqinv}.  Black dotted lines display the model presented in
    this work, WLOR relation as in Eq. \ref{eq_lobs} considering the values reported in Table \ref{tabResultsZ}.\label{figRichness4}}
\end{figure*}

We compute the cluster richness using the position of haloes and
subhaloes in the simulated field of view of 5 Mpc on the
side. Specifically, from each halo projection, we count the total
number of subhaloes and haloes in a cylinder with a radius equal to
$R_{200}$, with respect to the cluster centre and height equal to
$\Delta$H=$10$ Mpc that corresponds to roughly $\Delta z = 0.002 \times (1+z)$, which represents an optimistic case reachable with spectroscopic data
\citep{euclidredbook,jauzac21,caminha23,daddona24,wst_whitep}.
The richness of the simulated clusters is computed by considering a
minimum stellar mass cut mimicking the observational measure that
depends on the survey depth and the corresponding magnitude limit of
the data set, photometric properties, and stellar mass determination.

In Fig.~\ref{figSubsandhaloes}, we show the convergence map, haloes
and subhaloes in the $(x,y)$ plane of our test halo at redshift
$z=0.22$ from Fig. \ref{figMaps}. Black circles show the location of
cluster subhaloes, the red-filled triangles all haloes in the cylinder
projected within $R_{200}$ including also the main halo. In this case,
we consider all (sub)haloes with $M_{\rm star}>0$. The magenta crosses
mark the haloes in a cylindrical ring between 2 and 3.5 times
$R_{200}$. These are chosen far enough from the cluster to be
representative of the halo projected density in the field.  This
sample is used to estimate the number of contaminant haloes
by subtracting the local background to correct the cluster richness
\citep{andreon12}, which we define as:
\begin{equation}
\lambda_{\rm obs}(M_{\rm star}>M_{\rm star,\,min}) = n_{\rm Subhaloes}
+ n_{\rm Haloes} - \dfrac{4}{33}n_{\rm Outer \, Haloes} \,.
\label{eqrichness}
\end{equation}
The geometrical factor 4/33 has been computed by rescaling the density
of haloes in the cylindrical ring between 2 and 3.5 $R_{200}$ to
account for the difference in area.  This richness estimate is
computed individually for each cluster projection, while for each
cluster, the intrinsic richness is calculated from $\lambda_{\rm int}
= n_{\rm Subhaloes}$+1, by counting the number of satellite galaxies in subhaloes
within $R_{200}$, plus the central.

As reference cases, in Fig. \ref{figRichness4}, we display the
weak-lensing mass versus the richness for clusters at $z=0.22$
simulated using the \texttt{GadgetX} prescription.  In the four panels
we consider different stellar mass cuts, from $M_{\rm
  star,\,min}=10^{10}\,h^{-1}M_{\odot}$ to $M_{\rm
  star,\,min}=10^{10.75}\,h^{-1}M_{\odot}$ with a step
$\mathrm{d}\log\left(M_{\rm star,\,min}\right)=0.25$.  For each
$M_{\rm star,\,min}$ case, we display three different sets of data
points. The first one is the true mass-intrinsic richness relation
that considers the true $M_{200}$ and all subhaloes within $R_{200}$
(magenta triangles, as in \citealt{chen24}); in this case, there are
no projection effects when counting the member galaxies, while bias
and uncertainties on the cluster mass are set to zero.  Second,
combining all three cluster projections, the green circles show
the case where we consider all satellite galaxies in a sphere of
radius $R_{200}$, plus the central, and adopt as cluster mass
$M_{wl}$, the one computed modelling the weak-lensing simulated
data. Third, the blue-filled circles, with the corresponding error
bars, show the observed richness with the corresponding Poisson
uncertainty and the weak-lensing mass $M_{wl}$, calculated as
described in Section \ref{subsec:2.3}.  For comparison, in the top two
panels, the magenta dashed lines show the results by \citet{chen24},
who modelled the intrinsic cluster richness as a function of the true
mass $M_{200}$, calibrated up to $M_{\rm
  star,\,min}\simeq10^{10}\,h^{-1}M_{\odot}$, using the same
simulations of this work.  Note that this model already fails to
describe the magenta triangle in the top-right panel for which $M_{\rm
  star,min}=10^{12.5}\,h^{-1}M_{\odot}$.  In all panels of Figure
\ref{figRichness4}, we show two log-log relations between mass and
richness described here calculated for
different minimum stellar mass cuts used in computing the richness.
Notice that the scatter increases
at higher stellar mass cuts and that the inverse relation (displayed
with a dot-dashed orange line) differs from the weak-lensing
mass-richness relation when the scatter and the errors in the richness
are larger. Notice that considering that our cluster sample is a true mass-selected at $z=0$, those results are not strongly affected by Malmquist-Eddington biases. For redshift $z>0$, the low mass sample is modulated by the stochasticity of the mass accretion history \citep{vandenbosch02,giocoli07,giocoli12b} and by the noisy estimate of the mass via weak lensing simulations, and the corresponding uncertainties. 

\begin{itemize}[itemsep=12pt]
    \item[$\bullet$] \textbf{Weak-lensing mass-observed richness
      relation} 
\end{itemize}

The solid blue lines display the best result of the linear regression
model:
\begin{equation}
\langle \log \left( \lambda_{\rm obs}(M_{\rm star}>M_{\rm star,\,min})
| M_{wl} \right) \rangle = A + B \log \left( \dfrac{M_{wl}}{M_{\rm
    p}}\right)\,, \label{eq_lobs}
\end{equation}
where masses are rescaled with respect to the pivot value $M_{\rm p} =
3\times 10^{14}\,h^{-1}M_{\odot}$, the intercept $A$ and the slope $B$
are computed using an MCMC Bayesian inference model, assuming a
Gaussian likelihood. The latter, accounting for error bars on both
axes is written as:
\begin{equation}
\mathcal{L} \propto \exp \left\{ - \dfrac{1}{2} \sum_i \left[\dfrac{
    \log\left(\lambda_{{\rm obs},i}\right) -\log
    \left(\lambda_i\right)}{ \sigma_i}\right]^2
\right\}\,,\label{eqRlike}
\end{equation}
with
\begin{equation}
 \sigma_i^2 = \sigma_{\log \lambda_{\mathrm{obs},i}}^2 + B^2
 \sigma_{\log M_{wl,i}}^2 \,, \label{eqRlike2}
\end{equation}
where $\sigma_{\log \lambda_{\mathrm{obs},i}}$ and $\sigma_{\log
  M_{wl,i}}$ are the uncertainties relative to the $i$-cluster. 
  Notice that we assume a Poisson uncertainty to the richness with $\sigma_{\log \lambda_{\mathrm{obs},i}} = \left(\ln(10) \sqrt{\lambda_i}\right)^{-1}$. 
We assume wide uniform priors for $A$ and $B$: $A \in [-50,50]$ and $B\in
[-1;5]$.  The cyan-shaded regions in Fig. \ref{figRichness4} display
the 1$\sigma$ uncertainty of the linear regression model parameters.
The observed richness is related to the
weak-lensing mass by a power-law relation $\lambda_{\rm obs}\propto
M_{wl}^{B}$, and considering that both observables have their
corresponding associated uncertainties, it follows that $ M_{wl}
\cancel{\propto} \lambda_{\rm obs}^{1/B}$
\citep{andreon10,andreon12,andreon12b,andreon13,andreon16}. \\

\begin{itemize}[itemsep=12pt]
    \item[$\bullet$] \textbf{Observed richness-weak-lensing mass
      relation} 
\end{itemize}
The orange dash-dotted lines in Fig.\ref{figRichness4}, display the
$inverse$ of the median posteriors of the linear regression:
\begin{equation}
\left\langle \log \left( \dfrac{M_{wl}}{M_{\rm p}} \right)
\right\rangle = C + D \log \left( \lambda_{\rm obs}(M_{\rm
  star}>M_{\rm star,\,min}) | M_{wl} \right)\,,\label{eqinv}
\end{equation}
that relates the weak-lensing mass to the observed richness by the
power-law relation $M_{wl}\propto \lambda_{\rm obs}^{D}$, underlining
that $D \neq 1/B$. For likelihood modelling of the linear regression
parameters $C$ and $D$, we use Equation \ref{eqRlike} substituting the richness with the mass, the associated uncertainty propagation can be read as
\begin{equation}
 \sigma_i^2 = \sigma_{\log M_{wl,i}}^2 + D^2 \sigma_{\log
   \lambda_{\mathrm{obs},i}}^2 \,. \label{eqRlike3}
\end{equation}

\begin{figure*}
   \centering
   \includegraphics[width=0.49\hsize]{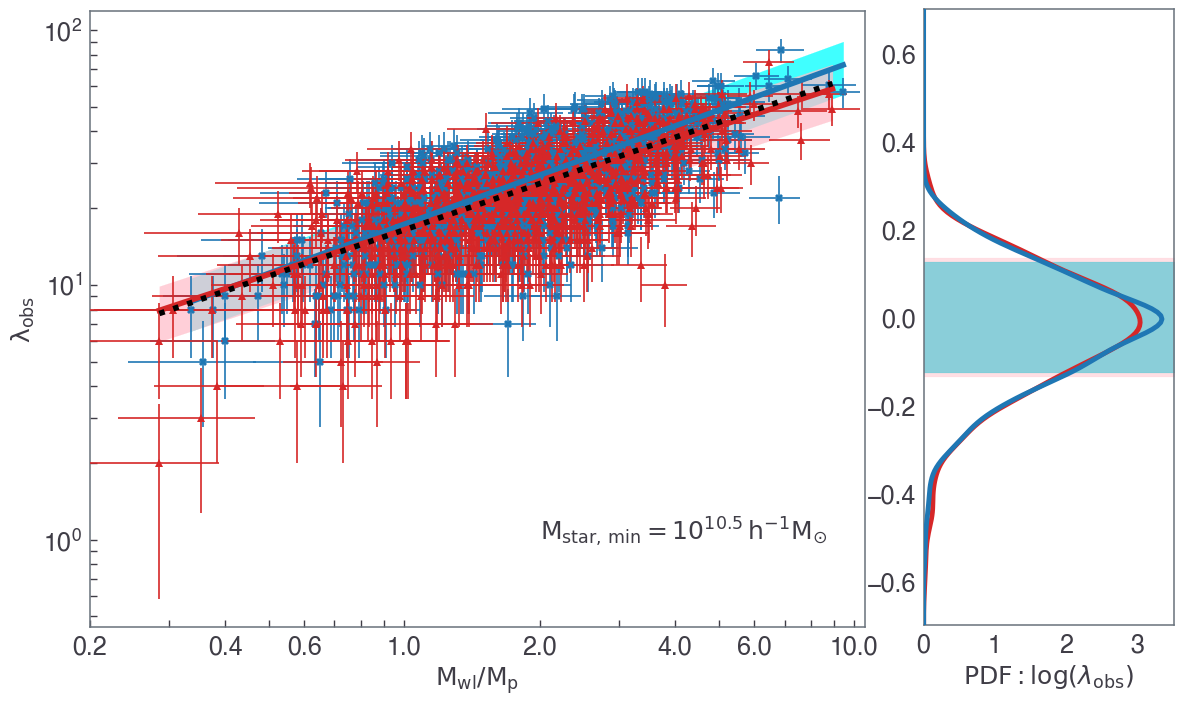}   
   \includegraphics[width=0.49\hsize]{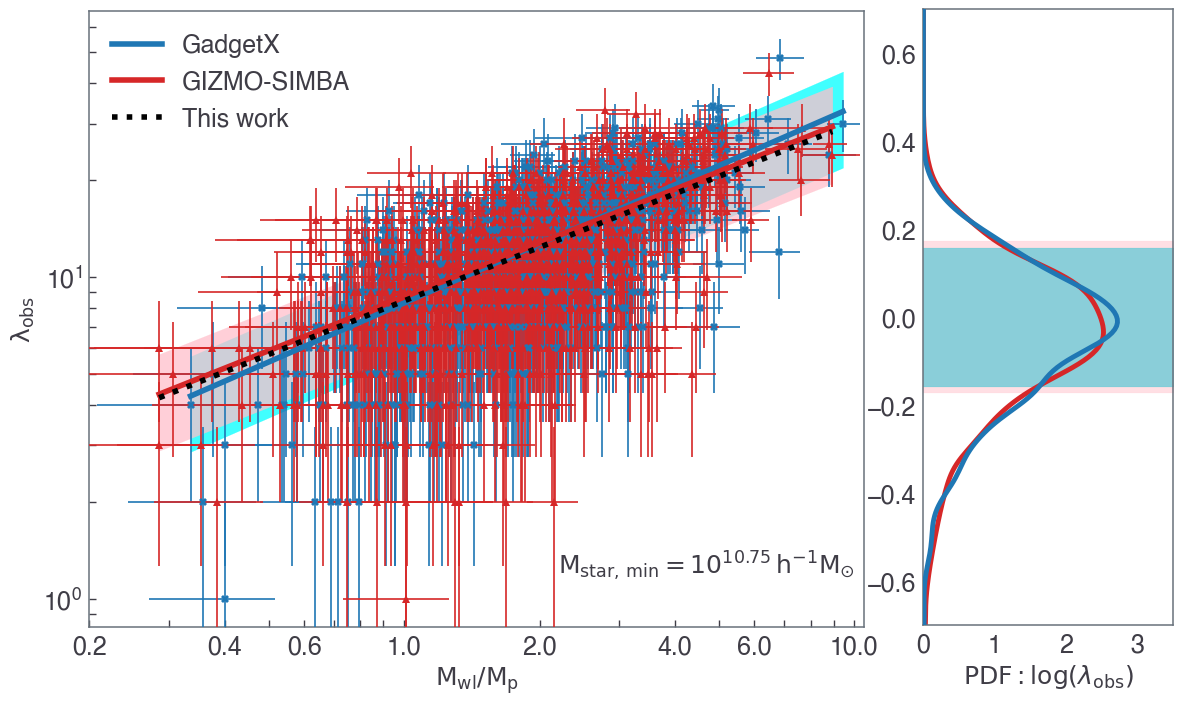}
   \caption{\label{figRichnessSims}Weak-lensing mass-observed richness
     relation measured for the clusters at redshift $z=0.22$ run with
     both hydrocodes \texttt{GadgetX} in blue and \texttt{GIZMO-SIMBA}  
     in red, respectively. The blue points are thus the same from
     Figure \ref{figRichness4}. The solid lines show the linear
     regression model results, with the shaded region indicating the
     $1\,\sigma$ uncertainty on the intercept and slope
     parameters. In each sub-figure -- where we consider two different stellar 
     mass cuts when computing $\lambda_{rm obs}$, the right sub-panel shows the PDf of 
     the scatter of the observed richness (using a Gaussian KDE) with respect to the best-fit linear regression model; the shaded bands mark the corresponding $1\,\sigma$ deviation: $\sigma_{\log \lambda}$.}
    \end{figure*}

While Figure \ref{figRichness4} refers only to \texttt{GadgetX}, in
the two panels of Fig.\ref{figRichnessSims}, we show the weak-lensing mass-observed richness (WLOR) relation
in both hydro-runs, considering the two largest stellar mass cuts:
\texttt{GadgetX} in blue and \texttt{GIZMO-SIMBA} in red,
respectively. The results of the linear regression Bayesian inferred model are shown with the corresponding coloured solid lines and the
shaded 1$\sigma$ uncertainty.  The figure shows that the results from
both hydro runs are consistent within the shaded uncertainty region
derived from the independent propagated uncertainties in the MCMC
analyses of the two hydro runs. The right sub-panels show the PDF of
the scatter of the observed richness with respect to the corresponding
linear regression model at a fixed weak-lensing mass, with the shaded
region marking the corresponding scatter $\sigma_{\log \lambda_{\rm
    obs}}$.  It is worth underlining that, on average, the scatter in the \texttt{GIZMO-SIMBA} cluster simulations is marginally larger than
the one measured in \texttt{GadgetX}.
Generally, the parameters $A$ and $B$ of the linear regression model
(see Eq. \ref{eq_lobs}) and derived $\sigma_{\log \lambda_{\rm obs}}$ depend on the stellar mass cut $M_{\rm star,\,min}$, redshift $z$ and the
hydro simulation considered.

\subsection{Redshift evolution}

Our analysis aims to model the observed richness-weak-lensing mass
relations for clusters at different redshifts from $z=0.12$ up to
$z=0.94$. In order to do so, we model all the nine considered
simulation snapshots and analytically describe the redshift evolution
of the parameters describing this scaling relation, namely the
intercept, the slope and the scatter.

\begin{figure*}
   \centering
   \includegraphics[width=0.33\hsize]{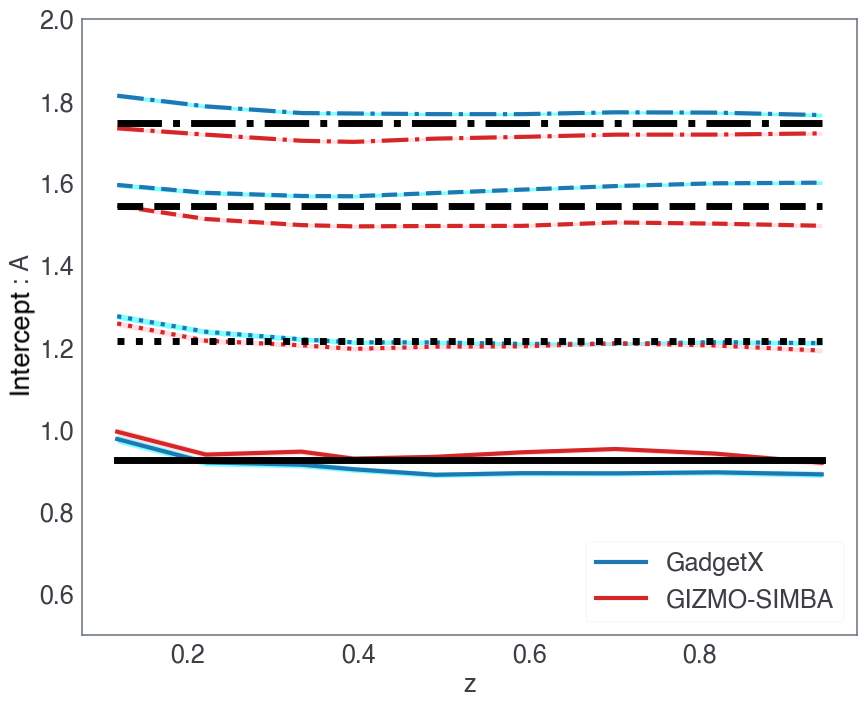}
   \includegraphics[width=0.33\hsize]{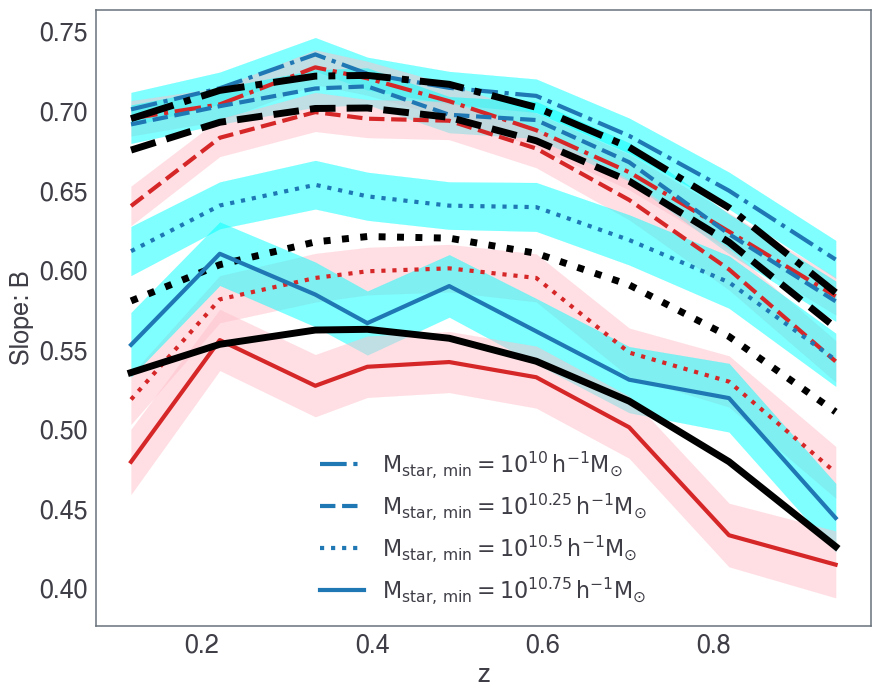}
   \includegraphics[width=0.33\hsize]{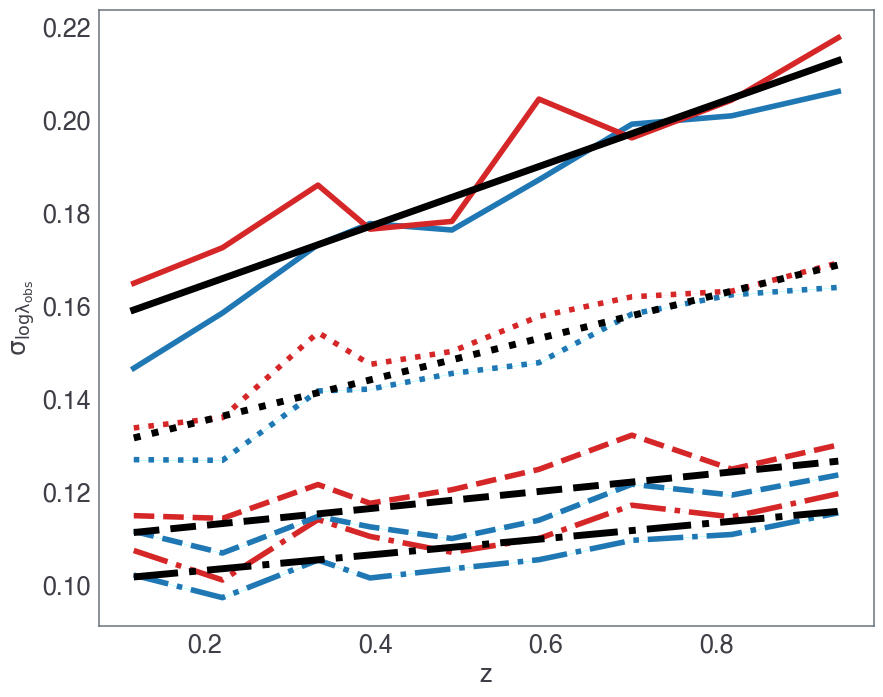}   
   \caption{\label{figParsZ} Intercept $A$, slope $B$, and scatter
     $\sigma_{\log \lambda_{\rm obs}}$ at a fixed weak-lensing cluster
     mass as a function of redshift. Blue and red curves refer to the
     two hydro simulations: \texttt{GadgetX} and \texttt{GIZMO-SIMBA},
     respectively. In all panels, the results referring to a given
     stellar mass cuts are reported with different line styles. The
     black curves show the best-fit models as a function of redshift,
     combining both hydro run results at a fixed stellar mass cut. The
     best-fit values are reported in Table~\ref{tabResultsZ}.}
 \end{figure*}

We summarise the results of the WLOR parameters in
Fig. \ref{figParsZ}, where they are displayed as a function of the
cluster redshifts, the four considered stellar mass cuts and both
hydro runs -- in blue and red for the \texttt{GadgetX} and
\texttt{GIZMO-SIMBA}, respectively.  We note that the intercept $A$ is
consistent with being redshift independent and increasing with the
stellar mass cut. \texttt{GadgetX} clusters have slightly larger
values than the \texttt{GIZMO-SIMBA} ones.  The slope parameter $B$
and the derived scatter $\sigma_{\log \lambda_{\rm obs}}$ instead manifest a moderate redshift dependence. A second-order polynomial $a+b\,z + c\, z^2$, fixing $c = -0.42$ independently of the stellar mass cut, describes quite well the slope, while the scatter is well fit by a
linear relation. In particular, the WLOR relations of the
\texttt{GadgetX} clusters have a steeper slope $B$ than the
\texttt{GIZMO-SIMBA} ones. Notice that the slope parameter is
relatively constant up to $z\simeq 0.55$, after which we are sensitive
to the variation and the cut-off of the stellar mass function
\citep{chen24} reflecting the complex interplay between galaxy
formation, environmental effects, and hierarchical structure
formation.  The black lines and curves in the three panels, styled
according to the different considered minimum stellar mass cuts
$M_{\rm star,\,min}$, represent the best-fit models combining both
hydro-runs. In Table~\ref{tabResultsZ}, we summarise the best-fit
parameters of redshift evolution and for different stellar mass cuts.
The dotted black lines -- labelled as 'This work' or as 'This work $M_{wl}$' -- reported in the Figs. \ref{figRichness4},
\ref{figRichnessSims} and \ref{figM200RichnessSims} have been computed
adopting the best-fit model parameters as reported in
Table~\ref{tabResultsZ}.  We follow the same procedure, combining the
results of both hydro simulations to model the redshift evolution of
the parameters of the observed richness-weak-lensing mass (ORWL) relation $C$, $D$ and the derived scatter $\sigma_{\log  M_{wl}}$, as reported in Table \ref{richnesstabResultsZ}.

It is worth underlining that we also retrieved the ORWL relation by adopting sharp richness cut, both in $\lambda_{\rm obs}$ and $\lambda_{\rm true}$, to further test that the Malmquist-Eddington biases do not affect our results. We used severe $\lambda_{\text{cut}} > [100, 60, 20, 10]$ for the four $M_{\rm star,\, min}$-cut samples, respectively, finding linear regressions slightly steeper but fully consistent with our reference results within the 1$\sigma$ confidence region. 

\begin{table*}[]
    \caption{Weak-lensing mass-observed richness relation parameters\label{tabResultsZ}.} \centering
    \begin{tabular}{l|c|c|c}
        $M_{\rm star,\,min}$         &  Intercept: $A$ & Slope: $B$ &  $\sigma_{\log \lambda_{\rm obs}}$\\ \hline 
        & & & \\
        $10^{10}\,M_{\odot}/h$      &        $1.74^{\pm0.03} $  & $0.664^{\pm 0.006} + 0.313^{\pm 0.010} z  - 0.42\, z^2$ & $0.100^{\pm 0.002} + 0.017^{\pm 0.004} z $ \\
        $10^{10.25}\,M_{\odot}/h$   &        $1.54^{\pm0.04} $  & $0.645^{\pm 0.006} + 0.310^{\pm 0.011} z - 0.42\, z^2$ & $0.109^{\pm 0.003} + 0.019^{\pm 0.004} z $ \\
        $10^{10.5}\,M_{\odot}/h$    &        $1.22^{\pm0.02} $  & $0.544^{\pm 0.008} + 0.361^{\pm 0.014} z - 0.42\, z^2$ & $0.126^{\pm 0.003} + 0.045^{\pm 0.005} z $ \\
        $10^{10.75}\,M_{\odot}/h$   &       $0.93^{\pm0.03} $  & $0.504^{\pm 0.010} + 0.313^{\pm 0.018} z - 0.42\, z^2$ & $0.152^{\pm 0.004} + 0.065^{\pm 0.007} z $ 
    \end{tabular}
    \tablefoot{Weak-lensing mass observed richness relation parameters as a
      function of redshift $z$ and considering different stellar mass
      cut $M_{\rm star,min}$, as shown in the left
      column.}
\end{table*}

\begin{figure}[!h]
   \centering
   \includegraphics[width=\hsize]{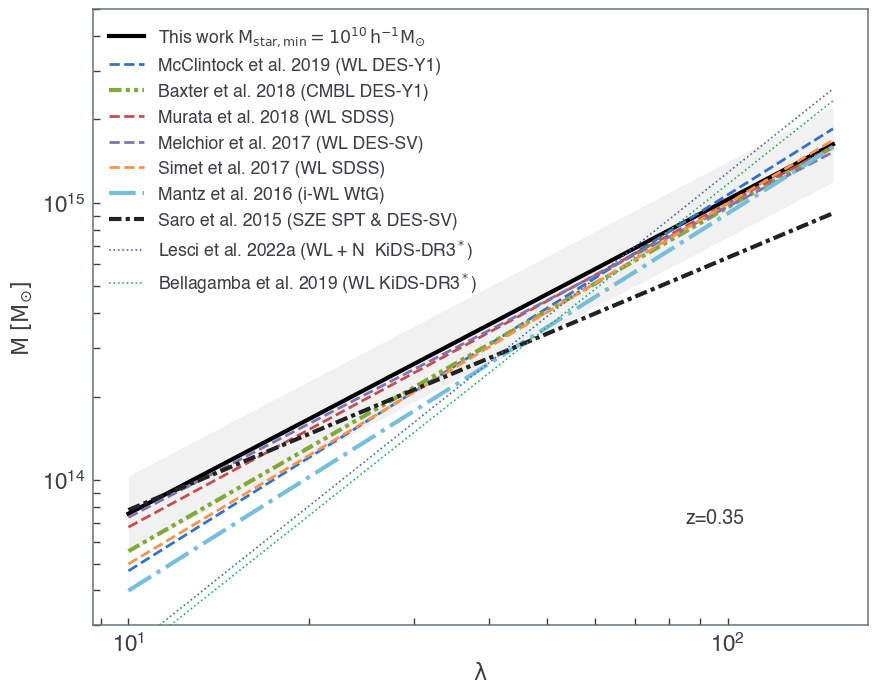}   
   \caption{Observed richness-weak-lensing mass relation comparison
     with different literature results. We compute our model
     considering an observed richness with a minimum stellar mass of
     $M_{\rm star,\,min}=10^{10}\,h^{-1}M_{\odot}$ and the parameters
     have been calculated at $z=0.35$. All redMaPPer \citep{rykoff14}
     richness definitions have been rescaled to DES-Y1 by
     \citep{mcclintock2019} (see their Tab. 5). The two dotted lines
     referring to KiDS-DR3 clusters from \citep{lesci22} and
     \citep{bellagamba19} have been computed with the AMICO cluster
     finder algorithm \citep{maturi19}.
   \label{figComparison}}
\end{figure}

\begin{table*}[]
    \caption{Observed richness-weak lensing mass relation parameters.\label{richnesstabResultsZ}}  \centering
    \begin{tabular}{l|c|c|c}
         $M_{\rm star,\,min}$ &  Intercept: $C$ & Slope: $D$ &  $\sigma_{\log M_{wl}}$\\ \hline 
         & & & \\
        $10^{10}\,M_{\odot}/h$     &  $-1.90^{\pm 0.05}$  & $1.12^{\pm 0.02} + 0.10^{\pm 0.08} z -0.18^{\pm 0.08} z^2$ & $0.126^{0.003} + 0.027^{0.006} z$ \\ 
        $10^{10.25}\,M_{\odot}/h$  &  $-1.62^{\pm 0.05}$  & $1.09^{\pm 0.02} + 0.08^{\pm 0.08} z -0.18^{\pm 0.08} z^2$ & $0.133^{0.004} + 0.032^{0.007} z $ \\
        $10^{10.5}\,M_{\odot}/h$   &  $-1.15^{\pm 0.04}$  & $1.04^{\pm 0.02} + 0.17^{\pm 0.09} z -0.40^{\pm 0.09} z^2$ & $0.155^{0.005} + 0.050^{0.009} z$ \\
        $10^{10.75}\,M_{\odot}/h$  &  $-0.73^{\pm 0.04}$  & $0.88^{\pm 0.02} + 0.37^{\pm 0.09} z -0.60^{\pm 0.08} z^2$ & $0.171^{0.006} + 0.073^{0.010} z$ 
    \end{tabular}
    \tablefoot{Observed richness weak-lensing mass relation parameters
      as a function of redshift $z$ and considering different stellar
      mass cut $M_{\rm star,min}$, as shown in the left
      column. The last column presents the
      redshift evolution of the logarithm of the weak-lensing mass at
      a given observed richness.}
\end{table*}

\subsection{Comparison with the literature}

In Fig. 10, we compare our ORWL relation -- considering $M_{\rm star,\,min}=10^{10}\,h^{-1}M_{\odot}$ -- with results from previous studies, rescaling various redMaPPer richnesses following Table 5 of \citet{mcclintock2019}. 

\citet{mcclintock2019} constrained the richness-mass scaling relation for galaxy clusters in DES Year 1 data using weak lensing, dividing clusters into richness ($\lambda \geq 20$) and redshift ($0.2 \leq z \leq 0.65$) bins. Mean masses were measured via stacked lensing signals, and their analysis, incorporating detailed systematic error considerations, provided precise constraints, demonstrating DES’s potential for cluster cosmology.

\citet{baxter18} used CMB lensing from SPT and DES Year 1 clusters (with mean redshift $z=0.45$), constraining the mass-richness relation with $\sim 17\%$ precision. Their analysis was primarily limited by statistical noise, with notable systematics from the thermal SZ effect and cluster miscentering. \citet{simet17} and \citet{murata18}
constrained the mass-richness relation for SDSS redMaPPer clusters in the redshift range $0.10 \leq z \leq 0.33$, though their methodologies differed: 
\citet{simet17}  applied an analytical model, while \citet{murata18} used forward modelling with a calibrated N-body emulator.

\citet{melchior17} analysed the richness-weak-lensing mass relation for over 8,000 redMaPPer clusters in DES Science Verification (SV) data, incorporating models that address systematic uncertainties. Their work extends the calibrated redshift range for redMaPPer clusters ($0.3 \leq z \leq 0.8$) while agreeing with prior weak-lensing calibrations. \citet{mantz16} constrained the scaling relation between redMaPPer richness and weak-lensing masses for an X-ray-selected sample from the Weighing the Giants project. \citet{saro15} cross-matched SPT-SZ and DES-SV clusters, confirming general consistency but identifying mild tensions in expected matching rates. They found that optical-SZE positional offsets followed a bimodal distribution.

\citet{bellagamba19} and \citet{lesci22} adopted AMICO, an alternative optical cluster finder that does not rely on colour-based selection, reducing biases from red-sequence detection. AMICO leverages galaxy luminosity, spatial distribution, and photo-$z$ data to provide probabilistic galaxy-cluster associations. Purity and completeness are evaluated using mock catalogues derived directly from observational data, avoiding reliance on numerical or semi-analytic models \citep{maturi19}. While
\citet{bellagamba19}  constrained the AMICO richness-mass relation using a stacked cluster sample, \citet{lesci22} included a model for the redshift evolution of the cluster mass function.

The redMaPPer algorithm identifies clusters using the red sequence, assuming a dominant population of red, passively evolving galaxies. In contrast, AMICO relies on galaxy luminosity, spatial distribution, and photometric redshifts, making it independent of galaxy colour. For richness estimation, redMaPPer uses a probability-based approach focused on red-sequence galaxies, whereas AMICO probabilistically incorporates all galaxies, considering spatial and magnitude constraints without colour dependence. This key distinction leads to differences in observed cluster populations, as shown in Fig. \ref{figComparison}.

From a numerical simulation perspective, selecting galaxies based solely on a stellar mass threshold better aligns with redMaPPer’s richness definition. This is represented by the solid black line in Fig. \ref{figComparison}, while the shaded region denotes 
$\sigma_{\log M}$ as modelled in The Three Hundred clusters (Table \ref{richnesstabResultsZ}). With a few exceptions, our simulated ORWL relation agrees well with observational data, particularly for richness $\lambda \geq 30$.

Before concluding this section, it is worth mentioning also the results from the CODEX clusters \citep{finoguenov20}. \citet{phriksee20}, for the low redshift sample, found a slope equal to unity, while \citet{kiiveri21} recovered a shallower slope in the mass-richness plane than \citet{mcclintock2019}.

\section{True mass-richness relations}
\label{sec_3}

The observed richness could also be directly used as a true mass
proxy.  In order to calibrate the relation with the true mass of the
cluster, it is worth underlining that for each cluster, we have three
observed richness estimates, depending on the projection we are
looking at, and one true mass $M_{200}$ defined as the spherical mass
within the radius $R_{200}$ which encloses 200 times the critical
comoving density of the universe $\rho_{\rm c}(z)$ at the considered
redshift. We model the linear regression between $M_{200}$ and
$\lambda_{\rm obs}$, and at different redshifts, using the same
methodology of the previous section.  In our MCMC analysis, we adopt
the same likelihood as in Eqs. \ref{eqRlike} and \ref{eqRlike2},
neglecting the uncertainty on the mass and considering only the
Poissonian error on the observed richness.

\begin{figure*}[!h]
   \centering
   \includegraphics[width=0.49\hsize]{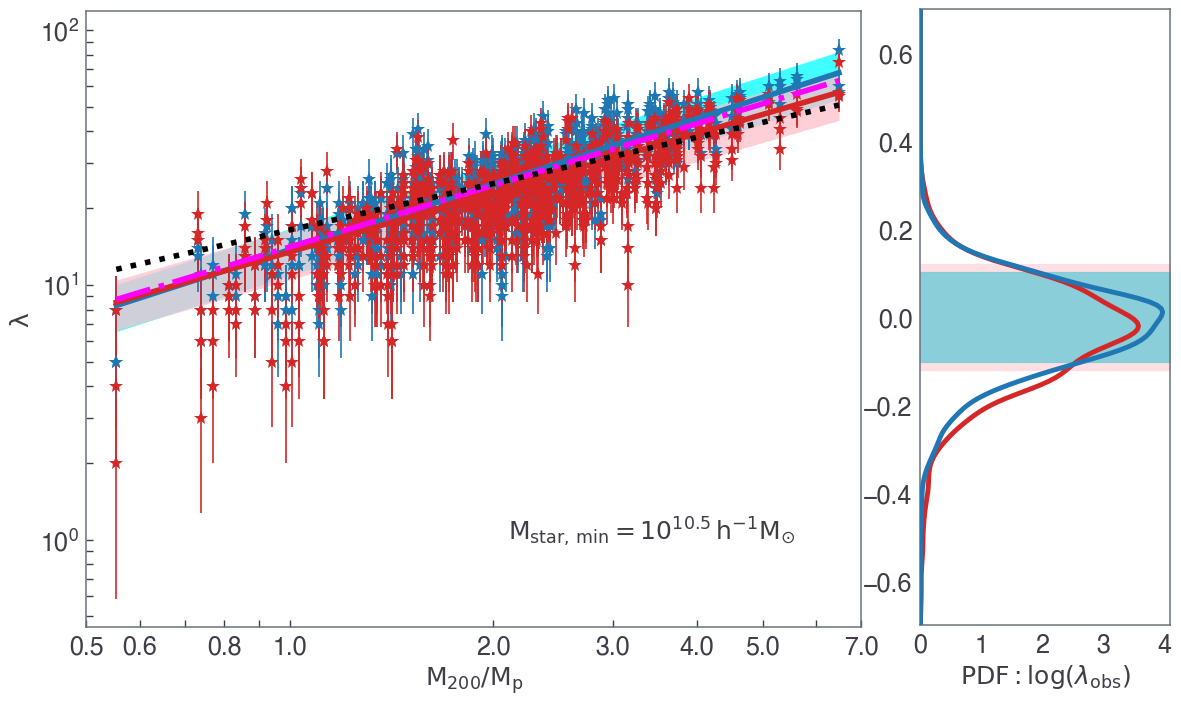}   
   \includegraphics[width=0.49\hsize]{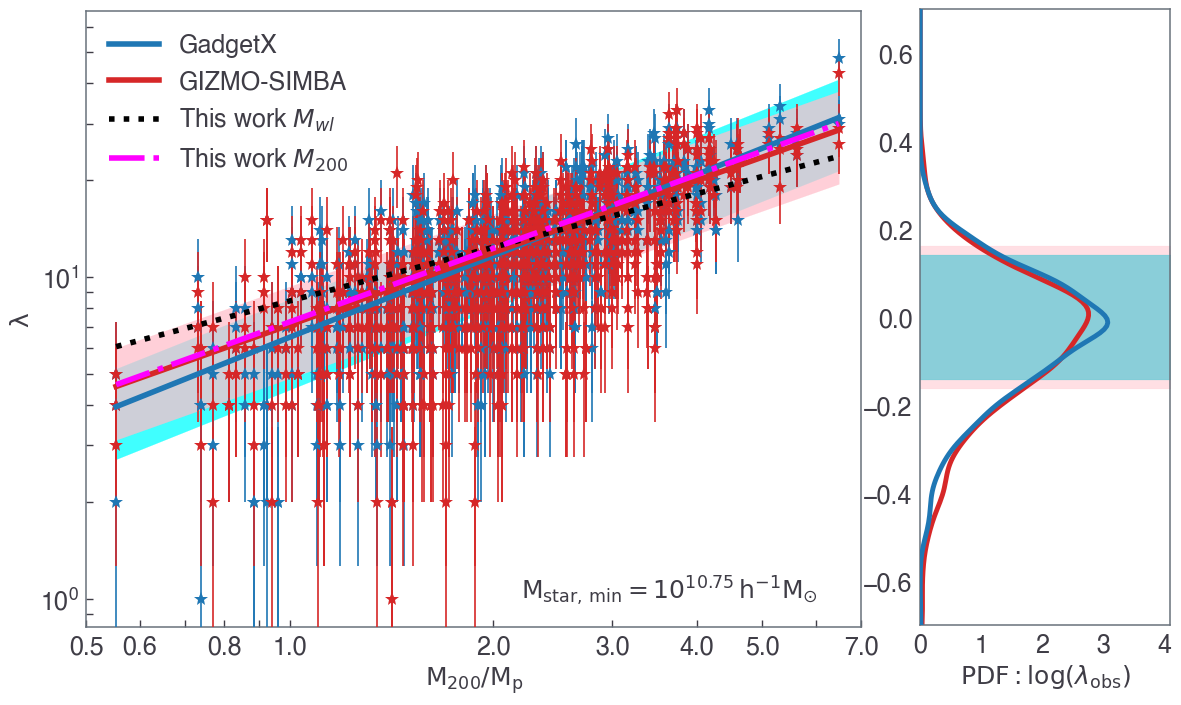}
   \caption{\label{figM200RichnessSims} True mass-observed richness
     relation for the clusters at redshift $z=0.22$ as measured in
     both hydro-runs. The solid lines show the linear regression model
     results, with the shaded region indicating the $1\,\sigma$
     uncertainty on the intercept and slope parameters. In each sub-figure, the right
     sub-panel shows the probability distribution function of the
     $\lambda_{\rm obs}$ with respect to the corresponding best-fit
     linear model, with the shaded regions marking the corresponding
     scatter $\sigma_{\log \lambda_{\rm obs}}$. The dotted black lines
     exhibit the model for the richness weak-lensing mass relation, as
     described in the last section.}
\end{figure*}

In Fig. \ref{figM200RichnessSims}, we display the true mass-observed
richness relation for clusters at $z=0.22$ for both hydro-runs and the
two highest stellar mass cuts. In both panels, the blue and red solid
lines display the best-fit linear regression models computed from the
median parameters of the posterior distributions.  Since, in this
case, we have the uncertainty only on the $y$-axis, we can directly
invert $\lambda_{\rm obs}\propto M_{200}^B$ into $M_{200} \propto
\lambda_{\rm obs}^{1/B}$. The black dotted lines in the figure
represent the linear regression models adopting the best-fit
parameters as in Table \ref{tabResultsZ}, as computed for the observed
richness-weak-lensing mass relation. The dot-dashed magenta lines
represent the linear regression model for the observed richness-true
mass case, the parameters of which are summarised in Table
~\ref{M200tabResultsZ}, obtained by combining the results of both
hydro-runs. From the figure, we observe that, on average, the
difference between the black dotted and magenta dot-dashed lines
reflects the weak-lensing mass bias as described in detail by
\citet{giocoli24}: the weak-lensing mass of low mass clusters is, on
average, more biased low than that of high mass systems with respect
to the true cluster mass. Notice that the difference between the
black-dotted and the dot-dashed magenta is also guided by the
dispersion of the observed richnesses.

\begin{table*}[]
    \caption{True mass-richness relation parameters. \label{M200tabResultsZ}} \centering
    \begin{tabular}{l|c|c|c}
         $M_{\rm star,\,min}$ &  Intercept: $A$ & Slope: $B$ &  $\sigma_{\log \lambda_{\rm obs}}$\\ \hline 
         & & & \\
        $10^{10}\,M_{\odot}/h$      &        $1.68^{\pm0.03} $  & $0.847^{\pm 0.005} + 0.206^{\pm 0.008} z - 0.18\,z^2$ & $0.062^{\pm 0.002} + 0.032^{\pm 0.004} z $ \\
        $10^{10.25}\,M_{\odot}/h$ &        $1.48^{\pm0.05} $  & $0.855^{\pm 0.006} + 0.161^{\pm 0.011} z - 0.18\, z^2$ & $0.077^{\pm 0.003} + 0.031^{\pm 0.003} z $ \\
        $10^{10.5}\,M_{\odot}/h$   &        $1.15^{\pm0.01} $  & $0.780^{\pm 0.009} + 0.145^{\pm 0.016} z - 0.18\, z^2$ & $0.104^{\pm 0.004} + 0.056^{\pm 0.006} z $ \\
        $10^{10.75}\,M_{\odot}/h$  &       $0.86^{\pm0.03} $  & $0.757^{\pm 0.012} + 0.049^{\pm 0.022} z - 0.18\, z^2$ & $0.136^{\pm 0.005} + 0.074^{\pm 0.008} z $ 
    \end{tabular}
    \tablefoot{True mass $M_{200}$-richness relation parameters as a
      function of redshift $z$ and considering different stellar mass
      cut $M_{\rm star,min}$, as shown in the left
      column.}
\end{table*}

As previously discussed, we model the evolution of the intercept $A$,
the slope $B$ and the derived scatter $\sigma_{\log \lambda_{\rm obs}}$ as a function of redshift and for the four considered stellar mass cuts, in
the true mass-observed richness relation.  We also note that, in this
case, the intercept is redshift-independent, the slope has a
second-order polynomial dependence on $z$ but with the second-order
term coefficient $c=-0.18$ and the scatter of the observed richness at
a given true mass $M_{200}$ depends linearly with redshift. It is
worth underlining that the scatter of the observed richness at a fixed
true mass is smaller than that of a fixed weak-lensing mass at a given
redshift and stellar mass cut, as expected being $M_{wl}$ a noisy proxy 
of the true underlying mass. 

\section{Summary and conclusions}
\label{sec_sc}

Cosmological numerical simulations play an important role in guiding
cluster cosmology studies.  Weak gravitational lensing is the leading
method in reconstructing galaxy clusters' projected matter density
distribution. However, derived weak-lensing masses are biased low with
respect to the true tridimensional mass, which is used in the halo
mass function models.

Using state-of-the-art hydrodynamical simulations, in this work, we
quantify the average weak-lensing bias dependences of two
hydro-solvers \texttt{GadgetX} and \texttt{GIZMO-SIMBA} and compare
with respect to a \texttt{DM-only} one
\citep{tinker08,sheth99b,despali16}.  Using cluster random
projections, we also calibrate the weak-lensing mass-richness relation
and its inverse and compare our findings with different literature
results. In what follows, we summarise our main results:
\begin{itemize}

\item All weak-lensing masses are, on average, negatively biased with
  respect to the corresponding true mass
  \citep{meneghetti07b,meneghetti10a,meneghetti10b,lee18}. While the ratio between the weak-lensing masses of \texttt{GadgetX} clusters
  and the three-dimensional \texttt{DM-only} ones are a few per cent
  higher, the ones derived from \texttt{GIZMO-SIMBA} are lower
  probably due to the shallower inner slope of the total density
  profile due to the strong AGN feedback \citep{meneghetti23}; when WL
  masses are rescaled with respect to the corresponding three-dimensional
  ones all runs provide the same answer for the weak lensing mass bias
  and its scatter and their redshift evolution.
  
\item Using halo and subhalo projected positions, we derive the
  observed richness accounting for local background contaminants.
  
\item The weak-lensing mass-observed richness relations derived from
  both hydrodynamical simulations are consistent within 1$\sigma$
  because of the propagated uncertainties; accordingly, we average the
  linear regression parameters from these simulations in our final
  model.
  
\item While the intercept parameter is redshift-independent and varies
  with the minimum stellar mass cut used to define the cluster
  richness, the slope, almost constant up to redshift $z=0.55$,
  changes with redshift following a second-order polynomial.
  
\item The redshift evolution of the scatter of the observed richness
  (weak-lensing mass) at a fixed weak-lensing mass (observed richness)
  linearly increases with redshift, and the stellar mass cut $M_{\rm
    star,\,min}$.
  
\item Our model for observed richness-weak-lensing mass is in good
  agreement with different literature results based on SDSS redMaPPer
  clusters when considering a minimum stellar mass cut $M_{\rm
    star,\,min}=10^{10}\,h^{-1}M_{\odot}$.
  
\item In our last section, we derive the linear regression parameters
  and their dependence on $z$ and $M_{\rm star,\,min}$ for the true
  mass-observed richness relation; we find that the scatter of
  $\lambda_{\rm obs}$ at a given true mass is smaller than the scatter
  at a given weak-lensing mass.
  
\end{itemize}
   
Numerical simulations have been extensively used to model the cluster
mass function as a function of redshift and overdensity, along with
their dependence on cosmology. However, the three-dimensional
overdensity mass of clusters is not directly observable, and its
reconstruction requires high-quality data and calibration with
advanced numerical simulations. By combining this with other mass
proxies obtained from multi-band observations, well-calibrated
mass-observable relations can be developed to estimate galaxy cluster
masses for larger samples with known observable properties. To
conclude, it is important to underline that establishing and
quantifying the relationship between weak-lensing cluster mass,
richness and potential systematics from baryonic physics is crucial
for advancing precision cluster cosmology studies.

\begin{acknowledgements}
LM and CG acknowledge the financial contribution from the PRIN-MUR
2022 20227RNLY3 grant 'The concordance cosmological model:
stress-tests with galaxy clusters' supported by Next Generation EU and
from the grant ASI n. 2024-10-HH.0 “Attività scientifiche per la
missione Euclid – fase E”.  \\ GC thanks the support from INAF theory
Grant 2022: Illuminating Dark Matter using Weak Lensing by Cluster
Satellites. \\ GD acknowledges the funding by the European Union -
NextGenerationEU, in the framework of the HPC project – “National
Centre for HPC, Big Data and Quantum Computing” (PNRR - M4C2 - I1.4 -
CN00000013 – CUP J33C22001170001). \\SB is supported by the Fondazione
ICSC, Spoke 3 Astrophysics and Cosmos Observations. National Recovery
and Resilience Plan (Piano Nazionale di Ripresa e Resilienza, PNRR)
Project ID CN00000013 "Italian Research Center on High-Performance
Computing, Big Data and Quantum Computing" funded by MUR Missione 4
Componente 2 Investimento 1.4: Potenziamento strutture di ricerca e
creazione di "campioni nazionali di R\&S (M4C2-19 )" - Next Generation
EU (NGEU); the National Recovery and Resilience Plan (NRRP), Mission
4, Component 2, Investment 1.1, Call for tender No. 1409 published on
14.9.2022 by the Italian Ministry of University and Research (MUR),
funded by the European Union – NextGenerationEU– Project Title
"Space-based cosmology with Euclid: the role of High-Performance
Computing" – CUP J53D23019100001 - Grant Assignment Decree No. 962
adopted on 30/06/2023 by the Italian Ministry of Ministry of
University and Research (MUR).  \\GC, GD, MM, LM, SB and FM are also
supported by the INFN InDark Grant.\\ 
This research was supported in part by grant NSF PHY-2309135 to the Kavli Institute for Theoretical Physics (KITP). \\ The authors acknowledge The Red
Española de Supercomputación for granting computing time for running
most of the simulations of The Three Hundred galaxy cluster project in
the Marenostrum supercomputer at the Barcelona Supercomputing Center.
\\WC and GY would like to thank Ministerio de Ciencia e Innovación for
financial support under project grant PID2021-122603NB-C21. WC is also
supported by the STFC AGP Grant ST/V000594/1 and the Atracci\'{o}n de
Talento Contract no. 2020-T1/TIC-19882 granted by the Comunidad de
Madrid in Spain. He also thanks the ERC: HORIZON-TMA-MSCA-SE for
supporting the LACEGAL-III project with grant number 101086388 and the
China Manned Space Project for its research grants.
This work has been made possible by the ’The Three Hundred’ collaboration.\footnote{https://www.the300-project.org}.
\end{acknowledgements}

\bibliography{aanda}
\newpage
\begin{appendix} 
\onecolumn
\section{Dark matter-only true mass-richness relations}

In this appendix, we report the results of the relations between the true
 \texttt{DM-only} masses and the observed richnesses as computed 
in both hydro runs. This is motivated by the large availability of dark matter-only cosmological simulations that are used as a baseline to construct mass proxy scaling relations based on semi-analytical, halo occupation distribution and baryon painting methods.

\begin{figure*}[!h]
   \centering
   \includegraphics[width=0.46\hsize]{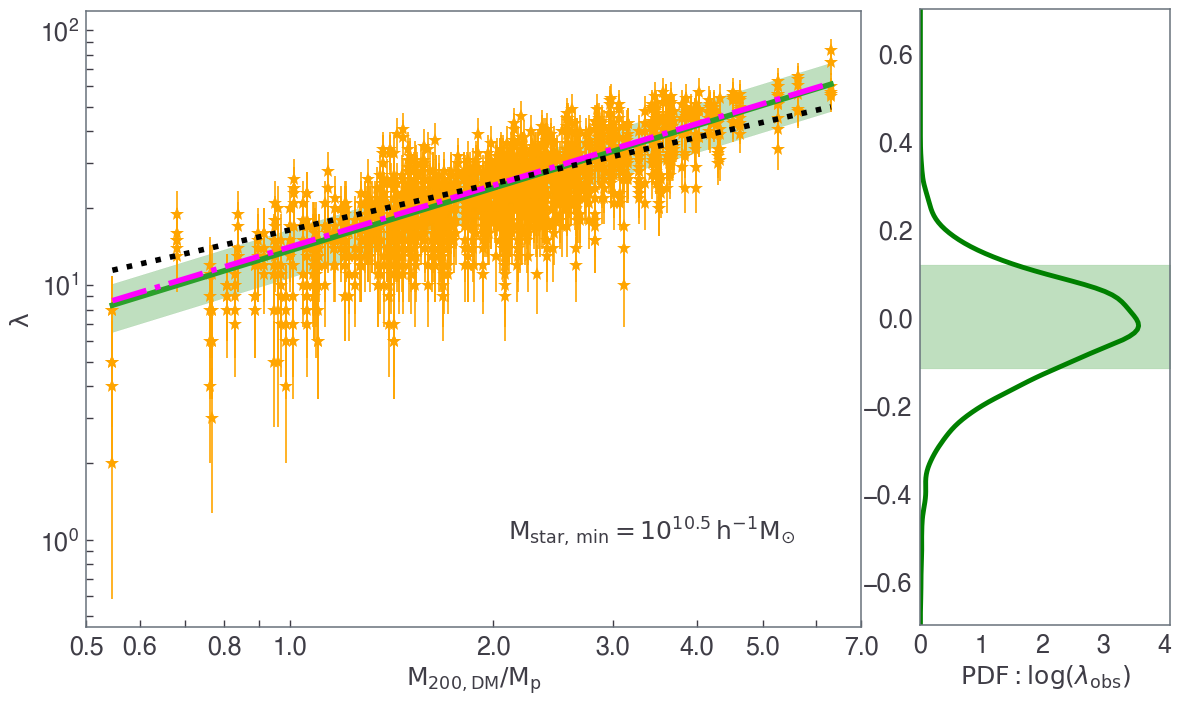}   
   \includegraphics[width=0.46\hsize]{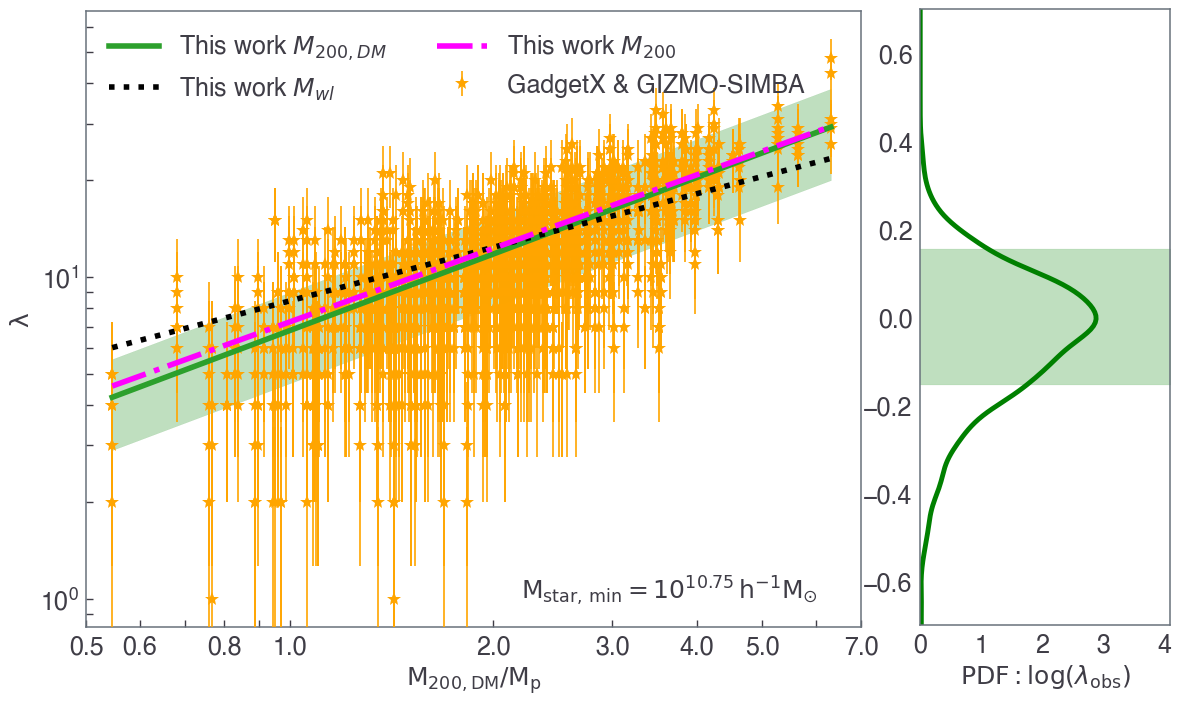}
   \caption{\label{afigM200DM} \texttt{DM-only} true mass-observed richness
     relation for the clusters at redshift $z=0.22$. The richnesses have been measured 
     using the two corresponding hydro runs -- \texttt{GadgetX} and \texttt{GIZMO-SIMBA} -- and considering the three projections per cluster.  The solid green lines show the linear regression model results, with the coloured corresponding shaded region indicating the $1\,\sigma$ uncertainty on the intercept and slope parameters. 
      The dotted black and magenta long-dashed lines report to the model for the true hydro and weak lensing masses, as discussed in the main text. The corresponding right sub-panels display the PDF of the observed richnesses with respect to the best-fit linear model, with the shaded regions marking the corresponding $\sigma_{\log \lambda_{\rm obs}}$.}
\end{figure*}

In Fig.~\ref{afigM200DM}, we show the \texttt{DM-only} true mass-observed richness relations for clusters at $z=0.22$, considering the two highest stellar 
mass cuts. The orange data points refer to the results for both hydro runs, considering 
three projections per cluster when computing the observed richnesses. The solid green lines show the linear regression best-fit model to the data.  In the figures, we also report the models based on the true corresponding hydro (long-dashed magenta) and derived weak lensing (dotted black) masses.

In Tab. \ref{M200DMtabResultsZ}, we summarise the evolution of the intercept $A$,
the slope $B$ and the derived scatter $\sigma_{\log \lambda_{\rm obs}}$ as a function of redshift computed for the four considered stellar mass cuts, in the \texttt{DM-only} true mass-observed richness relation.  

\begin{table*}[!h]
    \caption{True \texttt{DM-only} mass-richness relation parameters.\label{M200DMtabResultsZ}} \centering
    \begin{tabular}{l|c|c|c}
         $M_{\rm star,\,min}$ &  Intercept: $A$ & Slope: $B$ &  $\sigma_{\log \lambda_{\rm obs}}$\\ \hline 
         & & & \\
        $10^{10}\,M_{\odot}/h$      &        $1.68^{\pm0.01} $  & $0.847^{\pm 0.004} + 0.207^{\pm 0.008} z - 0.18\,z^2$ & $0.078^{\pm 0.002} + 0.025^{\pm 0.003} z $ \\
        $10^{10.25}\,M_{\odot}/h$ &        $1.48^{\pm0.02} $  & $0.854^{\pm 0.006} + 0.162^{\pm 0.010} z - 0.18\, z^2$ & $0.086^{\pm 0.001} + 0.043^{\pm 0.001} z $ \\
        $10^{10.5}\,M_{\odot}/h$   &        $1.15^{\pm0.01} $  & $0.782^{\pm 0.009} + 0.143^{\pm 0.016} z - 0.18\, z^2$ & $0.109^{\pm 0.002} + 0.053^{\pm 0.004} z $ \\
        $10^{10.75}\,M_{\odot}/h$  &       $0.86^{\pm0.01} $  & $0.756^{\pm 0.012} + 0.047^{\pm 0.022} z - 0.18\, z^2$ & $0.138^{\pm 0.003} + 0.076^{\pm 0.005} z $ 
    \end{tabular}
    \tablefoot{True \texttt{DM-only} mass $M_{200}$-richness relation parameters as a
      function of redshift $z$ and considering different stellar mass
      cut $M_{\rm star,min}$, as reported in the left
      column.}
\end{table*}

\end{appendix}

\end{document}